\documentclass[11pt,a4paper]{article}
\usepackage[T1]{fontenc}
\usepackage{lmodern}

\usepackage[a4paper,textwidth=0.72\paperwidth,textheight=0.76\paperheight,centering]{geometry}
\usepackage{amsmath,amssymb}
\usepackage{graphicx}
\usepackage[numbers,sort&compress]{natbib}
\usepackage{mathrsfs,microtype}
\usepackage{tikz}
\usetikzlibrary{arrows.meta,decorations.pathreplacing}
\usepackage[colorlinks=true,linkcolor=blue,citecolor=blue,urlcolor=blue]{hyperref}
\definecolor{figblue}{RGB}{28,70,140}
\definecolor{figred}{RGB}{150,30,30}
\newcommand{\dd}{\mathrm d}
\newcommand{\TF}{\mathrm{TF}}
\allowdisplaybreaks[1]
\numberwithin{equation}{section}

\hypersetup{pdftitle={Obstruction to local entropy laws on arbitrary black-hole cuts in five-dimensional Einstein--Gauss--Bonnet gravity},pdfauthor={Cendikiawan Suryaatmadja and Robert B. Mann}}

\title{Obstruction to local entropy laws on arbitrary black-hole cuts in five-dimensional Einstein--Gauss--Bonnet gravity}
\author{Cendikiawan Suryaatmadja$^{1}$ and Robert B.\ Mann$^{1,2}$\\[8pt]
{\small $^{1}$Department of Physics and Astronomy, University of Waterloo,}\\
{\small Waterloo, Ontario N2L 3G1, Canada}\\[2pt]
{\small $^{2}$Perimeter Institute, Waterloo, Ontario N2L 2Y5, Canada}\\[6pt]
{\small E-mail: \href{mailto:cendikiawan.suryaatmadja@uwaterloo.ca}{\texttt{cendikiawan.suryaatmadja@uwaterloo.ca}},
\href{mailto:rbmann@uwaterloo.ca}{\texttt{rbmann@uwaterloo.ca}}}}
\date{}

\begin{document}
\maketitle

\begin{abstract}
We study local entropy laws on arbitrary cuts of event horizons in five-dimensional Einstein--Gauss--Bonnet gravity, treated as an exact theory. No covariant local entropy can both reproduce the stationary black-hole entropy and be nondecreasing between every ordered pair of cuts, provided its density is smooth on an invariant open set of off-shell metric and cut jets containing the jet data of a flat planar cut and the data at every point of every cut used in the argument. No global bound on derivative order is assumed. The counterexamples are vacuum black holes perturbed by a gravitational wave, with a strongly hyperbolic formulation near the relevant horizon interval and its stationary future. Shear makes the intrinsic curvature of a cut depend on its bending. Stationary normalization fixes the coefficient of this curvature term, while local corrections cannot cancel its response to bending. The obstruction appears at second order in wave amplitude on a smooth horizon, without a merger or topology change.
\end{abstract}

\noindent\textit{Keywords}: Einstein--Gauss--Bonnet gravity, black-hole entropy, second law, event
horizons

\section{Introduction}\label{sec:introduction}

Black-hole entropy in general relativity is proportional to horizon area. Under the hypotheses of Hawking's area theorem, it is nondecreasing toward the future~\cite{Hawking1971}. On a smooth portion of an event horizon, this statement compares any two compact spacelike sections, or cuts, ordered along the null generators. It does not require a preferred slicing. One may advance a cut by a small amount while changing its slope and bending substantially. The area depends on the position of the cut along each generator, but not on derivatives of the function that describes the cut.

In higher-curvature gravity the stationary entropy contains additional geometric terms. In Einstein--Gauss--Bonnet (EGB) gravity, the Jacobson--Myers expression includes the intrinsic scalar curvature of the cut~\cite{JacobsonMyers1993}. Away from stationarity, this curvature depends on how the cut varies from one generator to the next. A law along a prescribed foliation therefore need not extend to every ordered pair of cuts. Agreement with the stationary entropy does not resolve the issue: terms involving the null expansions and shears can vanish on stationary cuts while changing the entropy during evolution.

Several second laws are known under perturbative or effective-field-theory assumptions. Wall established a second law for linear perturbations on arbitrary cuts~\cite{Wall2015}. Hollands, Kov\'acs and Reall constructed an entropy obeying a second law to quadratic order in perturbations within gravitational effective field theory~\cite{HollandsKovacsReall2022}. Davies and Reall subsequently obtained a law without an amplitude expansion within that effective description~\cite{DaviesReall2024}. The latter uses a horizon foliation satisfying derivative bounds and permits errors at the order of omitted terms. They also constructed an EGB entropy through quadratic order in the Gauss--Bonnet coupling within the effective description~\cite{DaviesReall2023}. These results leave open the existence of an exact law for unrestricted cuts when EGB gravity is treated as an exact theory.

Obstructions to broader entropy laws have appeared in other settings. Entropy can decrease in Lovelock black-hole mergers~\cite{Liko2008,SarkarWall2011}, and Gauss--Bonnet terms can violate the quantum focusing conjecture even in weakly curved spacetime~\cite{FuKoellerMarolf2017}. Neither result rules out every local entropy on smooth event-horizon cuts of fixed topology. Bhattacharyya \textit{et al.}\ identified difficulties in constructing a covariant entropy form for arbitrary Lovelock horizon sections~\cite{BhattacharyyaEtAl2017}. Hollands, Wald and Zhang discussed rapidly varying cuts and the restrictions imposed by effective field theory~\cite{HollandsWaldZhang2024}.

Here we prove an obstruction in five-dimensional EGB gravity. We consider covariant local entropies built from the metric, the embedded cut and the null direction tangent to the horizon. Their densities must be smooth on a specified open set of off-shell metric and embedding jets, including the jets of a flat planar cut and those at every point of every cut used in the argument. The derivative order need not be bounded globally. Within this class, stationary normalization is incompatible with monotonicity between every ordered pair of cuts of the vacuum horizons constructed below. These horizons have compact three-sphere cuts, a smooth interval with nonzero shear, and an exactly stationary future. A strongly hyperbolic formulation exists near the interval containing the shear and near the stationary future.

Monotonicity on every graph forces the local entropy rate to be independent of all derivatives of the cut function. This conclusion follows from locality and positivity, without a bound on the derivatives of the cuts. A large-radius limit at fixed physical coupling then isolates the degree-two curvature terms in any entropy with the stated regularity. The stationary value fixes the intrinsic-curvature coefficient, leaving two extrinsic-curvature coefficients to test. The shear of a vacuum horizon makes those necessary identities inconsistent: one correction is excluded at first order in wave amplitude, and the other cannot remove the entropy rate's dependence on the tracefree part of the cut Hessian at second order.

Section~\ref{supp:setup} specifies the geometric conventions, entropy class and result. Section~\ref{app:allcuts} derives the consequence of monotonicity on all cuts. Section~\ref{app:vacuum} constructs the exact vacuum family and establishes its horizon and regularity properties. Section~\ref{app:reduction} gives the local finite-order argument and the large-radius reduction. Section~\ref{sec:obstruction} evaluates the remaining coefficients and obtains an ordered pair with decreasing entropy. Section~\ref{sec:discussion} discusses the scope of the obstruction.

\section{Geometric setting and statement of the result}\label{supp:setup}

\subsection{Theory, horizons and null normals}

We use signature $(-++++)$, Newton's constant $G_5>0$, zero cosmological constant and Gauss--Bonnet coupling $\alpha_{(2)}>0$. The action is
\begin{equation}
 I=\frac{1}{16\pi G_5}\int\sqrt{-g}\,
 \bigl(\mathcal R+\alpha_{(2)}\mathcal L_{\rm GB}\bigr)\dd^5x,
 \qquad
 \mathcal L_{\rm GB}=\mathcal R^2-4\mathcal R_{ab}\mathcal R^{ab}
 +\mathcal R_{abcd}\mathcal R^{abcd}.
 \label{set:action}
\end{equation}
We write $\tilde\alpha=4\alpha_{(2)}$ and work on the branch connected continuously to general relativity.

The event horizon is the metric causal boundary
$\mathcal H=\partial J^-(\mathscr I^+_{\rm R})$ associated with one selected asymptotically flat end. In EGB gravity it need not be a causal boundary for every gravitational polarization~\cite{ReallTanahashiWay2014}. A cut $\mathcal C$ is a smooth compact spacelike section meeting each generator once. We write $\mathcal C_1\preceq\mathcal C_2$ when the second cut lies nowhere to the past of the first along the generators. The comparison is made along each generator separately.

The cut metric, derivative, scalar curvature and measure are $q_{AB}$, $D_A$, $R_q$ and $\dd\mu_q$. Choose null normals with $\ell_+$ future directed and tangent to the horizon, normalized by
\begin{equation}
 g(\ell_+,\ell_-)=-1,
 \qquad \ell_\pm\longmapsto e^{\pm b}\ell_\pm.
 \label{set:normals}
\end{equation}
The smooth function $b$ changes the normalization of the normals, not their directions. For cut tangent vectors $e_A$, define
\begin{equation}
\begin{aligned}
 \chi^\pm_{AB}&=g(\nabla_{e_A}\ell_\pm,e_B),
 &\theta_\pm&=q^{AB}\chi^\pm_{AB},\\
 \omega_A&=-g(\ell_-,\nabla_{e_A}\ell_+),
 &\nabla_{\ell_+}\ell_+&=\kappa\ell_+.
\end{aligned}
\label{supp:chi}
\end{equation}
The tracefree part of $\chi^+$ is the horizon shear, $\theta_+$ is its expansion, and $\omega_A$ is the normal connection. Cut indices are raised using $q$, and $\TF$ denotes the tracefree part. We use $\sqrt q=\sqrt{\det q_{AB}}$ and $q^{ab}=q^{AB}e_A^ae_B^b$ for the inverse cut metric lifted to spacetime.

The boost transformations follow directly from the product rule and $g(\ell_\pm,e_A)=0$:
\begin{equation}
\begin{aligned}
 (\chi^\pm_{AB})'
 &=e^{\pm b}\bigl[\chi^\pm_{AB}
       \pm(\partial_A b)g(\ell_\pm,e_B)\bigr]
   =e^{\pm b}\chi^\pm_{AB},\\
 \theta_\pm'&=e^{\pm b}\theta_\pm,
 \qquad \omega_A'=\omega_A+\partial_A b.
\end{aligned}
\label{set:boost}
\end{equation}
If $b$ is extended along the horizon, the nonaffinity transforms as
\begin{equation}
 \kappa'=e^b\bigl(\kappa+\ell_+(b)\bigr).
 \label{set:kappaboost}
\end{equation}
For a boost constant along each generator, this becomes $\kappa'=e^b\kappa$. A component has boost weight $s$ when a constant boost multiplies it by $e^{sb}$. In particular, $\chi^+$ and $\chi^-$ have opposite weights, so $\theta_+\theta_-$ and $\chi^+_{AB}\chi^{-AB}$ are invariant.

\subsection{Entropy class and stationary normalization}\label{set:class}

We rescale the physical entropy $\mathcal S$ by $S=4G_5\mathcal S$ and write $\mathcal A=\int_{\mathcal C}\dd\mu_q$. The Jacobson--Myers functional is
\begin{equation}
 S_{\rm JM}[\mathcal C]
 =\int_{\mathcal C}(1+2\alpha_{(2)}R_q)\dd\mu_q.
 \label{set:JM}
\end{equation}
On a round stationary cut of radius $L$, $R_q=6/L^2$ and the area is $A_3L^3$, where $A_3=2\pi^2$ is the area of the unit three-sphere. The required stationary value is therefore
\begin{equation}
 S_{\rm stat}(L)=A_3\bigl(L^3+12\alpha_{(2)}L\bigr).
 \label{set:stat}
\end{equation}

An admissible entropy is the integral of a covariant local density whose only data are the spacetime metric, the embedded cut and the null direction tangent to the horizon. No additional clock, preferred foliation, matter field or normalization of the null normals is supplied. The integrated functional is invariant under local boosts. A representative density may change by a cut divergence, provided that transformation law is an identity throughout its domain.

We impose the following regularity on this domain. A jet records the values and derivatives of the metric and cut embedding at a point. The density is defined on a diffeomorphism- and boost-invariant open set of off-shell metric and cut jets of all orders. This set contains the flat planar jet in Minkowski spacetime. It also contains the metric and embedding jets at every point of every cut of the stationary horizons, including their bifurcation cuts, and of every cut in the vacuum horizon intervals constructed below. No bounds are imposed on the derivatives of the cut. At the flat planar jet, spacetime and intrinsic curvatures and both null second fundamental forms vanish.

The topology is the product topology in independent jet coordinates for the metric and cut together. Each coordinate representative of the density is continuous, and all its iterated directional derivatives exist and are jointly continuous in the jet and direction arguments. Section~\ref{app:localorder} proves that this smoothness implies finite derivative order in a neighborhood of each jet. The order need not be fixed on the whole domain.

These are off-shell assumptions: the metric and embedding can be varied independently near the specified data, even when the neighboring data do not satisfy the field equations. The physical horizons on which monotonicity is tested do satisfy the vacuum EGB equations. Smoothness at the planar jet excludes expressions singular when curvature and second fundamental forms vanish, including unregularized ratios of vanishing invariants. It allows a finite Taylor expansion without requiring analyticity or a polynomial entropy. The planar limit is a limit of local data, not a change in cut topology.

We ask whether an entropy in this class can satisfy both of the following conditions:
\begin{enumerate}
\item[(i)] It equals \eqref{set:stat} on every round cut of every sufficiently large stationary Boulware--Deser black hole~\cite{BoulwareDeser1985}.
\item[(ii)] On every vacuum horizon in the family below, it obeys
\begin{equation}
 S[\mathcal C_1]\le S[\mathcal C_2]
 \quad\hbox{whenever}\quad
 \mathcal C_1\preceq\mathcal C_2.
 \label{main:law}
\end{equation}
\end{enumerate}
Condition (i) fixes values at stationarity, while condition (ii) concerns all cuts during evolution. We do not assume that stationary normalization uniquely determines the nonstationary functional.

\subsection{The vacuum family and the obstruction}\label{set:family}

The construction in section~\ref{app:vacuum} gives exact vacuum metrics
\begin{equation}
 g_{L,\epsilon}=L^2\bar g_{\lambda,\epsilon},
 \qquad \lambda=\frac{\tilde\alpha}{L^2},
 \qquad 0\le\lambda\le\lambda_0,
 \qquad |\epsilon|<\epsilon_0.
 \label{red:family}
\end{equation}
The radius $L$ sets the physical length scale, and $\epsilon$ is the wave amplitude. The properties needed in the proof are as follows.

At $\epsilon=0$, the metrics are stationary Boulware--Deser metrics with horizon radius $L$. In a future neighborhood of the horizon, every perturbed member agrees with a stationary Boulware--Deser metric of the same horizon radius. At $\lambda=0$, the dimensionless metrics $\bar g_{0,\epsilon}$ are Ricci flat. Parameter-dependent diffeomorphisms identify the relevant horizon tubes with a fixed hypersurface with three-sphere sections; all parameter derivatives include this identification. Every member is invariant under a connected group of isometries that acts transitively on the three-sphere of horizon generators, and the identification is equivariant, so that the cuts of constant tube coordinate along the generators are homogeneous.

On the limiting Ricci-flat family, suppressing bars on cut quantities, we have
\begin{equation}
 \chi^+=\epsilon\Sigma+O(\epsilon^2),
 \qquad \theta_+=O(\epsilon^2),
 \qquad {\rm tr}_{q_0}\Sigma=0,
 \label{red:shear}
\end{equation}
where $q_0$ is the round background metric. The first-order shear $\Sigma$ is nonzero somewhere and vanishes on a final open interval. The absence of first-order expansion follows from the vacuum Raychaudhuri equation and the stationary future boundary condition. With $\ell_+=\partial_v$,
\begin{equation}
 \partial_v\theta_+-\kappa\theta_+
 =-\chi^+_{AB}\chi^{+AB}
 =-\epsilon^2\Sigma_{AB}\Sigma^{AB}+O(\epsilon^3).
 \label{red:raychaudhuri}
\end{equation}
The first-order expansion satisfies a homogeneous equation and vanishes on the final stationary interval, so it vanishes throughout the smooth horizon interval. The square of the first-order shear enters at the next order.

The coordinate derivatives of $\bar g_{\lambda,\epsilon}$ in a neighborhood of the tube, up to any prescribed finite order, are bounded uniformly and depend continuously on $\lambda$ down to $\lambda=0$ at fixed $\epsilon$, and the Ricci-flat family $\bar g_{0,\epsilon}$ is twice continuously differentiable in $\epsilon$ together with these derivatives. The constants $\lambda_0$ and $\epsilon_0$ may depend on that order. The horizon has no caustics, topology change or entering generators on the developed slab, and its stationary continuation has future-complete generators. The full modified harmonic formulation is strongly hyperbolic near the slab and stationary future, including for perturbations that break the symmetry used in the construction~\cite{KovacsReall2020}.

\paragraph{Result.}
No entropy in the class of section~\ref{set:class} satisfies both (i) and (ii). For each entropy in this class satisfying the stationary normalization (i), some member of the vacuum family admits two ordered smooth compact cuts on which the entropy decreases. The member of the family and the cuts may depend on the entropy.

The two parameters in \eqref{red:family} serve different purposes. The amplitude $\epsilon$ controls the shear, whereas increasing $L$ at fixed $\alpha_{(2)}$ makes spacetime curvature small compared with $\alpha_{(2)}^{-1}$. This scaling compares solutions of one fixed EGB theory. It does not impose a bound on cut derivatives. An arbitrarily weakly curved horizon still admits cuts whose graph functions vary rapidly across the generators.

The reduction in section~\ref{app:reduction} uses bounded sets of dimensionless graph data and only a finite Taylor expansion near the planar jet. The freedom to choose unrestricted cuts enters separately, through the exact consequence of condition (ii) proved in section~\ref{app:allcuts}. This separation is needed to pass from a perturbative contradiction to an ordered pair of cuts on which the exact entropy decreases.

\section{Cut geometry and monotonicity}\label{app:allcuts}

The requirement of monotonicity on every ordered pair of cuts imposes a local condition stronger than positivity along a chosen foliation. We first derive this condition without using the field equations. We then compute how the area and intrinsic-curvature integrals respond to the geometry of a cut.

\subsection{Graphs and local entropy variations}\label{cut:graphs}

Choose coordinates $(v,x^A)$ on a smooth horizon interval, with $\ell_+=\partial_v$ and the labels $x^A$ constant along each generator. A cut is the graph $v=\psi(x)$. Its tangent vectors differ from those of a constant-$v$ cut by $\psi_A\ell_+$, so nullness and orthogonality of the generator give
\begin{equation}
(q_\psi)_{AB}(x)=q_{AB}(\psi(x),x).
\label{cut:induced}
\end{equation}
The absence of slope terms means that every smooth graph with values in the interval is spacelike, however large its derivatives. Two cuts are ordered precisely when their graph functions are pointwise ordered. This comparison is made along each generator separately; it does not compare parameter values on different generators.

The freedom in the graph derivatives is local. For example, in coordinates centered at a point, the graph
\begin{equation}
\psi(x)=v_0+\frac12 X_{AB}x^Ax^B
\label{cut:quadraticgraph}
\end{equation}
has value $v_0$, zero slope and prescribed Hessian $X_{AB}$ there. Multiplying its quadratic part by a smooth cutoff equal to one near the point leaves these data unchanged. Shrinking the support keeps the graph inside the horizon interval for every fixed $X$, even when the eigenvalues of $X$ are large. The same construction with a finite Taylor polynomial prescribes any finite collection of graph derivatives. This changes the cut in a fixed spacetime and does not increase the spacetime curvature.

Let $f(x)$ be smooth. The deformation $\psi\mapsto\psi+t f$ moves the cut along the generators, and it is a forward deformation for $t>0$ when $f\ge0$. Local finite order and compactness, established in Sec.~\ref{app:localorder}, allow integration by parts in its first variation. Define the Euler--Lagrange density by
\begin{equation}
\left.\frac{\dd S[\psi+t f]}{\dd t}\right|_{t=0}
=\int\mathcal E[S]f\,\dd^3x.
\label{cut:euler}
\end{equation}
Here and below $\dd^3x$ denotes integration of densities on the manifold of generators. The density $\mathcal E[S]$ is smooth and local: if two graphs agree on an open set, their Euler--Lagrange densities agree there. Compactness and the absence of a boundary remove all boundary terms.

We use $\psi_A=D_A\psi$ and $\psi_{AB}=D_AD_B\psi$, with $D$ the covariant derivative of the constant-$v$ cut, evaluated at $v=\psi(x)$. At a point of zero slope, this Hessian agrees with the Hessian for the graph metric and with the coordinate second derivative. The response to a change of bending is
\begin{equation}
\mathcal E^{AB}[S]
=\left.\frac{\partial\mathcal E[S]}{\partial\psi_{AB}}\right|_{\psi_A=0},
\label{cut:hessianresponse}
\end{equation}
where the other graph jet coordinates are held fixed. Given a symmetric tensor $X_{AB}$, we denote the corresponding directional derivative under $\psi_{AB}\mapsto\psi_{AB}+sX_{AB}$ by $\delta_X$. Thus $\delta_X\mathcal E[S]=\mathcal E^{AB}[S]X_{AB}$ at zero slope. The tensor $X$ need not be positive: it prescribes a change in the local Hessian of the cut. The nonnegative function $f$ in Eq.~\eqref{cut:euler} separately specifies a forward motion from that cut.

\subsection{Consequences of monotonicity}\label{app:positivity}

Suppose $S$ is nondecreasing on every ordered pair of graphs with values in an interval $(v_1,v_2)$. For every smooth $f\ge0$, the graphs $\psi$ and $\psi+t f$ are ordered for sufficiently small $t>0$. Hence
\begin{equation}
0\le\lim_{t\to0^+}\frac{S[\psi+t f]-S[\psi]}{t}
=\int\mathcal E[S]f\,\dd^3x.
\label{cut:positivitytest}
\end{equation}
If $\mathcal E[S]$ were negative at a point, continuity would make it negative on a neighborhood, and a nonzero nonnegative $f$ supported there would contradict this inequality. It follows that $\mathcal E[S]\ge0$ on every graph. On the constant graph $\psi\equiv s$, define
\begin{equation}
e(x,s)=\left.\mathcal E[S]\right|_{\psi\equiv s}(x).
\label{cut:constantdensity}
\end{equation}
This is smooth in $x$ and $s$ and is nonnegative.

\paragraph{Comparison paths.}\label{app:minmax}

Positivity for all graphs also eliminates every dependence of $\mathcal E[S]$ on derivatives of $\psi$. To prove this, compare a graph with the constant cuts by two monotone paths obtained by rounding their pointwise minimum and maximum. Where a rounded path agrees locally with a constant cut, locality fixes its entropy rate. Positivity allows the remaining contributions to be discarded, giving bounds that coincide as the rounding is removed.

Fix $\psi$ and choose $c,d\in(v_1,v_2)$ with
$c<\min\psi\le\max\psi<d$. Let $\rho_\delta$ be a nonnegative even smooth function supported in $[-\delta,\delta]$, with integral one, where
$0<\delta<\min\{\min\psi-c,d-\max\psi\}$. For $c\le s\le d$, set
\begin{equation}
\begin{split}
\psi_s^-(x)&=\int_{\mathbb R}\rho_\delta(t)
                 \min\{s,\psi(x)+t\}\,\dd t,\\
\psi_s^+(x)&=\int_{\mathbb R}\rho_\delta(t)
                 \max\{s,\psi(x)+t\}\,\dd t.
\end{split}
\label{cut:smoothminmax}
\end{equation}
These are smooth jointly in $s$ and $x$, since
\begin{equation}
\partial_s\psi_s^-
=\int_{s-\psi(x)}^\infty\rho_\delta(t)\,\dd t,
\qquad
\partial_s\psi_s^+=1-\partial_s\psi_s^-.
\label{cut:pathrates}
\end{equation}
Both rates lie between zero and one. Evenness gives the endpoints exactly:
\begin{equation}
\psi_c^-=c,\qquad \psi_d^-=\psi,\qquad
\psi_c^+=\psi,\qquad \psi_d^+=d.
\label{cut:minmaxends}
\end{equation}
The two paths are therefore nondecreasing and stay within $(v_1,v_2)$.

On the open set $\{\psi>s+\delta\}$, the lower graph $\psi_s^-$ equals $s$ locally and has $s$ derivative one. Its Euler--Lagrange density there is $e(x,s)$; the contribution from the complement is nonnegative. Similarly, on $\{\psi<s-\delta\}$ the upper graph $\psi_s^+$ equals $s$ locally and has rate one. Consequently,
\begin{equation}
\begin{aligned}
\frac{\dd}{\dd s}S[\psi_s^-]
&\ge\int_{\{\psi>s+\delta\}}e(x,s)\,\dd^3x,\\
\frac{\dd}{\dd s}S[\psi_s^+]
&\ge\int_{\{\psi<s-\delta\}}e(x,s)\,\dd^3x.
\end{aligned}
\label{cut:pathbounds}
\end{equation}
Along constant graphs, the first-variation formula gives
\begin{equation}
S[d]-S[c]=\int\!\int_c^d e(x,s)\,\dd s\,\dd^3x.
\label{cut:constantpath}
\end{equation}
Integrating Eq.~\eqref{cut:pathbounds} from $c$ to $d$ and exchanging the order of integration yields
\begin{equation}
\begin{aligned}
S[\psi]-S[c]&\ge\int\!\int_c^{\psi(x)-\delta}e(x,s)\,\dd s\,\dd^3x,\\
S[d]-S[\psi]&\ge\int\!\int_{\psi(x)+\delta}^{d}e(x,s)\,\dd s\,\dd^3x.
\end{aligned}
\label{cut:integratedpaths}
\end{equation}
The first line is a lower bound. Subtracting the second from Eq.~\eqref{cut:constantpath} gives an upper bound:
\begin{equation}
\int\!\int_c^{\psi(x)-\delta}e(x,s)\,\dd s\,\dd^3x
\le S[\psi]-S[c]\le
\int\!\int_c^{\psi(x)+\delta}e(x,s)\,\dd s\,\dd^3x.
\label{cut:minmaxbounds}
\end{equation}
The difference between either bound and the integral with upper endpoint $\psi(x)$ is at most
\[
\delta\int\max_{s\in[c,d]}e(x,s)\,\dd^3x.
\]
Smoothness and compactness therefore imply
\begin{equation}
S[\psi]-S[c]=\int\!\int_c^{\psi(x)}e(x,s)\,\dd s\,\dd^3x.
\label{cut:localpotential}
\end{equation}
No bound on the derivatives of the rounded graphs entered this limit: their contributions outside the constant regions were discarded using positivity.

Keep $c,d$ fixed and replace $\psi$ by $\psi+t f$. Differentiation of the upper integration limit gives
\begin{equation}
\int\mathcal E[S]f\,\dd^3x
=\left.\frac{\dd}{\dd t}S[\psi+t f]\right|_{t=0}
=\int e(x,\psi(x))f(x)\,\dd^3x.
\label{cut:potentialvariation}
\end{equation}
Since this holds for every smooth $f$, including functions supported in arbitrarily small neighborhoods, the densities agree pointwise:
\begin{equation}
\mathcal E[S]=e(x,\psi(x))\ge0,
\qquad \mathcal E^{AB}[S]=0.
\label{cut:allcuts}
\end{equation}
In fact, the first equality removes all graph-derivative dependence, not only the dependence on the Hessian measured by the second equality.

\paragraph{Homogeneous horizons.}\label{app:homogeneous}

The horizons constructed in section~\ref{app:vacuum} are homogeneous: a connected group of isometries of the spacetime preserves the horizon, maps generators to generators, preserves each cut of constant $v$ and acts transitively on the three-sphere of generators. Covariance of the entropy gives $S[\psi\circ\phi]=S[\psi]$ for the induced maps $\phi$ of the three-sphere. The density $e(x,s)\,\dd^3x$ of Eq.~\eqref{cut:constantdensity} is therefore invariant, and equals $e_{\rm hom}(s)\,\dd\mu_{\Omega_3}$, where $\dd\mu_{\Omega_3}$ is the invariant volume element on the three-sphere of generators, normalized to total volume $A_3$. Write $S_{\rm hom}(s)$ for the entropy of the homogeneous cut $\psi\equiv s$. For constant graphs Eq.~\eqref{cut:localpotential} reads $S_{\rm hom}(s)-S_{\rm hom}(c)=A_3\int_c^se_{\rm hom}(s')\,\dd s'$, and for a general graph it becomes
\begin{equation}
S[\psi]=\frac{1}{A_3}\int_{S^3}S_{\rm hom}\bigl(\psi(x)\bigr)\,\dd\mu_{\Omega_3}(x).
\label{cut:homogeneous}
\end{equation}
An entropy that is nondecreasing on all ordered cuts of such a horizon is thus determined on every cut by its values on the homogeneous cuts. Equation~\eqref{cut:homogeneous} contains no derivative of $\psi$ and no derivative of the entropy. By Eq.~\eqref{cut:induced} the area satisfies it identically, on every horizon of this kind.

\subsection{From derivative dependence to entropy decrease}\label{app:pair}

The proof above uses monotonicity only to establish nonnegativity of the Euler--Lagrange density on every graph. Its contrapositive is useful: if that density depends on derivatives of $\psi$, it must be negative at some point of some graph. Choose a nonzero smooth $f\ge0$ supported where $\mathcal E[S]<0$. Smoothness then gives
\begin{equation}
S[\psi+t f]-S[\psi]
=t\int\mathcal E[S]f\,\dd^3x+O(t^2)<0
\label{cut:decrease}
\end{equation}
for sufficiently small $t>0$, although
$\mathcal C_\psi\preceq\mathcal C_{\psi+t f}$. Thus it suffices to show that, for an exact entropy with the required stationary value, $\mathcal E[S]$ cannot be independent of all spatial derivatives of $\psi$ on every horizon in the vacuum family. On the homogeneous horizons it equally suffices that Eq.~\eqref{cut:homogeneous} fails for some graph, because the derivation of that equation used monotonicity only through the nonnegativity of this density.

\subsection{Area and intrinsic-curvature variations}\label{app:area}

The constant-$v$ metrics evolve as
\begin{equation}
\partial_vq_{AB}=2\chi^+_{AB}.
\label{cut:qdot}
\end{equation}
Writing $\sqrt q=\sqrt{\det q_{AB}}$, the determinant identity gives
\begin{equation}
\partial_v\sqrt q
=\frac12\sqrt q\,q^{AB}\partial_vq_{AB}
=\theta_+\sqrt q.
\label{cut:volumeevolution}
\end{equation}
Equation~\eqref{cut:induced} implies that the area
$\mathcal A[\psi]=\int\sqrt q\,|_{v=\psi(x)}\,\dd^3x$ contains no graph derivatives. Its first variation is therefore
\begin{equation}
\mathcal E[\mathcal A]=\sqrt q\,\theta_+\big|_{v=\psi(x)}.
\label{cut:areaeuler}
\end{equation}
Under the same graph deformation, the induced metric varies according to
\begin{equation}
\left.\frac{\dd(q_{\psi+t f})_{AB}}{\dd t}\right|_{t=0}
=2f\chi^+_{AB}.
\label{cut:metricrate}
\end{equation}
The variation of the intrinsic Einstein--Hilbert integral is
$-\int(G_q)^{AB}\delta q_{AB}\,\dd\mu_q$, up to a cut divergence. Since the cut is compact without boundary, Eq.~\eqref{cut:metricrate} yields
\begin{equation}
\mathcal E\left[\int R_q\,\dd\mu_q\right]
=-2\sqrt q\,(G_q)^{AB}\chi^+_{AB}.
\label{cut:JMeuler}
\end{equation}
Unlike the area variation, this expression can depend on the bending of the cut through its intrinsic Einstein tensor.

\subsection{The curvature of a bent graph}\label{app:bending}

At zero slope, change the Hessian by $X_{AB}$ while holding the cut value and all other graph jet coordinates fixed. Equation~\eqref{cut:induced} shows that the graph metric and its first derivatives at the point are unchanged. Its second derivatives obey
\begin{equation}
\delta_X\partial_C\partial_D(q_\psi)_{AB}
=2\chi^+_{AB}X_{CD}.
\label{cut:qsecond}
\end{equation}
Indeed, at zero slope the chain rule gives
\begin{equation}
\partial_C\partial_D(q_\psi)_{AB}
=\left[\partial_C\partial_Dq_{AB}
 +(\partial_vq_{AB})\psi_{CD}\right]_{v=\psi(x)},
\label{cut:chainrule}
\end{equation}
where the derivatives of $q_{AB}$ on the right are taken at fixed $v$. Varying the Hessian changes only the second term, and Eq.~\eqref{cut:qdot} gives Eq.~\eqref{cut:qsecond}.

A metric change $h_{AB}$ that vanishes at a point together with its first derivatives changes the Ricci tensor there by
\begin{equation}
\delta R_{AB}=\frac12\left(
\partial_C\partial_Ah^C{}_B
+\partial_C\partial_Bh^C{}_A
-\partial^C\partial_Ch_{AB}
-\partial_A\partial_Bh^C{}_C\right),
\label{cut:linricci}
\end{equation}
with indices raised by the unperturbed metric. Substituting Eq.~\eqref{cut:qsecond} gives
\begin{equation}
\delta_X(R_q)_{AB}
=(\chi^+X+X\chi^+)_{AB}-\theta_+X_{AB}
-({\rm tr}_qX)\chi^+_{AB}.
\label{cut:ricci}
\end{equation}
Here ${\rm tr}_qX=q^{AB}X_{AB}$ and, for example,
$(\chi^+X)_{AB}=\chi^+_{AC}q^{CD}X_{DB}$.
This dependence vanishes on a stationary horizon, where $\chi^+=0$, but survives in the presence of shear. It is the geometric contribution that the entropy corrections must cancel if Eq.~\eqref{cut:allcuts} is to hold.

\section{Construction of the vacuum horizons}\label{app:vacuum}

The proof requires a vacuum horizon that shears on an earlier interval and is exactly stationary on a later one. It also requires a smooth limit as the dimensionless Gauss--Bonnet coupling tends to zero. We now construct the family described in Sec.~\ref{supp:setup}, including the bounds on coordinate derivatives and the dependence on the parameters used in the entropy reduction.

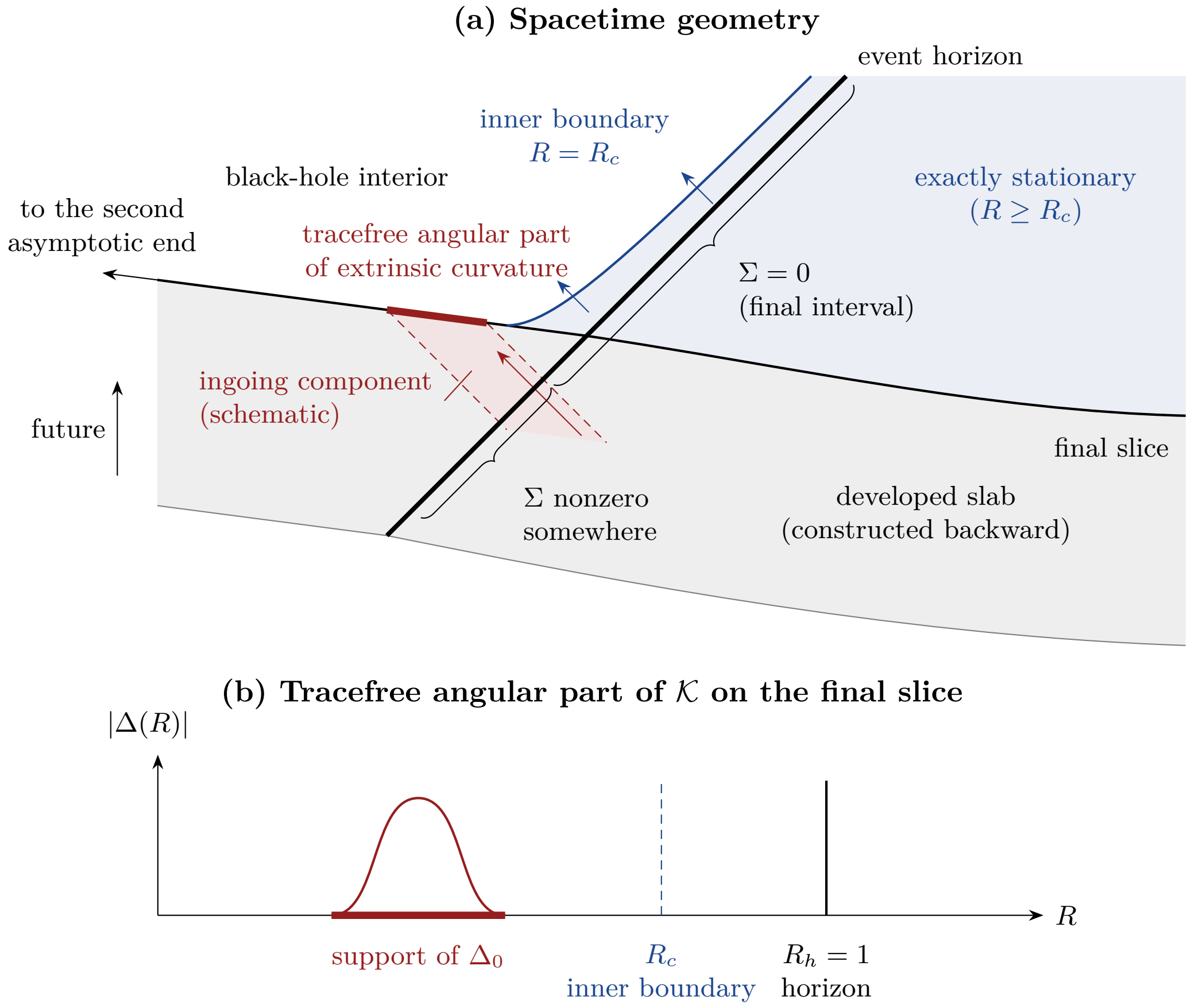
\begin{figure}[p]
\centering

\begin{tikzpicture}[
  x=1.15cm,
  y=1.15cm,
  >=Stealth,
  font=\footnotesize
]
  \node[font=\small\bfseries] at (4.5,7.15)
    {(a) Spacetime geometry};

  \fill[gray!13]
    (-0.3,4.56) -- (4,4)
    .. controls (6,3.7) and (8,3.25) .. (10,3.2)
    -- (10,0.9)
    .. controls (7,1.0) and (4,1.6) .. (2,2)
    -- (-0.3,2.30) -- cycle;

  \fill[figblue!9]
    (3.2,4.104186) -- (4,4)
    .. controls (6,3.7) and (8,3.25) .. (10,3.2)
    -- (10,6.6) -- (6.25,6.6)
    .. controls (4.5,4.875) and (3.7,4.104186)
    .. (3.2,4.104186)
    -- cycle;

  \fill[figred!12]
    (2,4.260465) -- (3,4.130233)
    -- (4.2,2.930233) -- (3.2,3.060465)
    -- cycle;

  \draw[figred,densely dashed]
    (2,4.260465) -- (3.2,3.060465);

  \draw[figred,densely dashed]
    (3,4.130233) -- (4.2,2.930233);

  \draw[->,figred]
    (3.95,3.0) -- (3.1,3.85);

  \draw[thick]
    (-0.3,4.56) -- (4,4)
    .. controls (6,3.7) and (8,3.25) .. (10,3.2);

  \draw[gray]
    (-0.3,2.30) -- (2,2)
    .. controls (4,1.6) and (7,1.0) .. (10,0.9);

  \draw[->]
    (-0.3,4.56) -- (-0.85,4.63);

  \draw[line width=1.5pt]
    (2,2) -- (6.6,6.6);

  \draw[figblue,thick]
    (3.2,4.104186)
    .. controls (3.7,4.104186) and (4.5,4.875)
    .. (6.25,6.6)
    coordinate[pos=0.35] (RcArrowA)
    coordinate[pos=0.75] (RcArrowB);

  \draw[->,figblue]
    (RcArrowA) ++(0.16,-0.16) -- ++(-0.32,0.32);

  \draw[->,figblue]
    (RcArrowB) ++(0.16,-0.16) -- ++(-0.32,0.32);

  \draw[figred,line width=2.4pt]
    (2,4.260465) -- (3,4.130233);

  \draw[
    decorate,
    decoration={brace,amplitude=4pt,mirror}
  ]
    (2.34,2.18) -- (3.64,3.48);

  \draw[
    decorate,
    decoration={brace,amplitude=5pt,mirror}
  ]
    (3.67,3.51) -- (6.68,6.52);

  \draw[->]
    (-0.7,2.6) -- (-0.7,3.55)
    node[midway,left] {future};

  \node[anchor=south,align=center,figred]
    at (2.5,4.48)
    {tracefree angular part\\of extrinsic curvature};

  \node[anchor=west,align=left,figred]
    (ingoing) at (0,3.35)
    {ingoing component\\(schematic)};

  \draw[figred,thin]
    (ingoing.east) -- (2.85,3.65);

  \node[anchor=north east] at (9.95,3.1)
    {final slice};

  \node[align=center] at (7.4,2.2)
    {developed slab\\(constructed backward)};

  \node[anchor=south west] at (6.6,6.6)
    {event horizon};

  \node[anchor=south east,align=center,figblue]
    at (4.95,5.6)
    {inner boundary\\$R=R_c$};

  \node[align=center,figblue] at (8.4,5.4)
    {exactly stationary\\($R\ge R_c$)};

  \node[align=center] at (1.5,5.6)
    {black-hole interior};

  \node[anchor=north west,align=left] at (3.25,2.6)
    {$\Sigma$ nonzero\\somewhere};

  \node[anchor=north west,align=left] at (5.4,4.85)
    {$\Sigma=0$\\(final interval)};

  \node[anchor=south,align=center] at (-0.85,4.68)
    {to the second\\asymptotic end};
\end{tikzpicture}

\par\medskip

\begin{tikzpicture}[
  x=1cm,
  y=1cm,
  >=Stealth,
  font=\footnotesize
]
  \node[font=\small\bfseries] at (5,2.55)
    {(b) Tracefree angular part of $\mathcal K$ on the final slice};

  \draw[->]
    (0,0) -- (10.2,0)
    node[right] {$R$};

  \draw[->]
    (0,0) -- (0,1.85);

  \node[anchor=south east] at (0.5,1.9)
    {$|\Delta(R)|$};

  \draw[figred,thick]
    (2,0)
    .. controls (2.55,0) and (2.45,1.35) .. (3,1.35)
    .. controls (3.55,1.35) and (3.45,0) .. (4,0);

  \draw[figred,line width=2.4pt]
    (2,0) -- (4,0);

  \node[anchor=north,align=center,figred] at (3,-0.2)
    {support of $\Delta_0$};

  \draw[figblue,densely dashed]
    (5.8,0) -- (5.8,1.55);

  \node[anchor=north,align=center,figblue] at (5.8,-0.2)
    {$R_c$\\inner boundary};

  \draw[thick]
    (7.7,0) -- (7.7,1.55);

  \node[anchor=north,align=center] at (7.7,-0.2)
    {$R_h=1$\\horizon};
\end{tikzpicture}

\caption{The vacuum construction, with spacetime position and perturbation
amplitude shown separately.
(a) Schematic time--radial diagram: time runs upward, radial light rays
run at $45^\circ$, and each point represents a homogeneous three-sphere.
The thick red segment marks the support of the tracefree angular part of
the extrinsic curvature on the final slice. The red shading and dashed edges
sketch one ingoing component at earlier times, with its forward propagation
across the event horizon indicated by the red arrow. Only this component
of the disturbance is shown. The gray slab is constructed by backward
evolution from
the final data. The first-order shear $\Sigma$ is nonzero somewhere on
the earlier bracketed interval and zero on the later interval.
The future region $R\ge R_c$ is exactly stationary; the blue arrows show
future characteristics crossing its inner boundary toward smaller $R$
(Sec.~\ref{app:vachorizon}). The final slice extends to a second
asymptotically flat end.
(b) Schematic magnitude $|\Delta(R)|=|\epsilon\Delta_0(R)|$ of the
coefficient specifying the tracefree angular part of $\mathcal K$ on the final slice, plotted against
the radial collar coordinate in units of the final horizon radius, so that
$R_h=1$. Its support lies strictly below $R_c<R_h$. The panels use separate
axes: the vertical coordinate is time in (a) and amplitude in (b).}
\label{fig:vacfamily}
\end{figure}

We solve the exact constraints on a complete spacelike slice that crosses the final horizon, prescribing a tracefree angular perturbation of the extrinsic curvature strictly inside it. Backward development through a finite slab produces nonzero first-order shear on an earlier horizon segment. We also identify a stationary future region that the interior disturbance cannot influence and trace its horizon backward (Figure~\ref{fig:vacfamily}). In this way the stationary future is prescribed as part of the solution, rather than obtained from a late-time decay theorem.

Beginning with Eq.~\eqref{vac:bd}, lengths are in units of $L$, so that the final horizon has areal radius $R=1$ and the coupling is $\lambda=\tilde\alpha/L^2$. We take $0\le\lambda\le\lambda_0$ and $|\epsilon|<\epsilon_0$. The constants may depend on the finite number of derivatives required by a given entropy. The dimensional family is recovered in Sec.~\ref{app:vacscaling}.

The final slice is chosen after the disturbance has crossed the horizon interval under consideration. Data inside the final horizon can therefore produce a perturbation on an earlier horizon segment under backward evolution. The gap between the support of the data and the final horizon cut supplies an undisturbed late segment. The direction of evolution is essential: no perturbation is being sent outward across the horizon in the physical future.

\subsection{The stationary exterior}
\label{app:spherical}

In a spherically symmetric spacetime, $R$ denotes the areal radius of the round three-spheres, and

\begin{equation}
F=1-(\nabla R)^2,\qquad A=1-F,\qquad K=1+\frac{\tilde\alpha F}{R^2}.
\label{set:FAK}
\end{equation}

The mass parameter $\mu$ and the generalized Misner--Sharp mass $\mathcal M$ of Refs.~\cite{MaedaNozawa2008,Suryaatmadja2026} are

\begin{equation}
\mu=R^2F+\tfrac12\tilde\alpha F^2,\qquad
\mathcal M=\frac{3\pi\mu}{8G_5}.
\label{set:mass}
\end{equation}

The parameter $\mu$ is constant in spherically symmetric vacuum regions, and at an asymptotically flat end $\mathcal M$ equals the ADM mass. The quantities $F$, $A$, $K$ and $\mu$ are used only where the geometry is spherically symmetric.

On the GR branch, which in spherical regions is the branch with $K>0$, the Boulware--Deser black hole~\cite{BoulwareDeser1985} reads

\begin{equation}
\begin{aligned}
\dd s^2&=-A\,\dd t^2+A^{-1}\dd R^2+R^2\dd\Omega_3^2\\
F&=\frac{2\mu}{R^2+(R^4+2\tilde\alpha\mu)^{1/2}}\qquad
R^2K=(R^4+2\tilde\alpha\mu)^{1/2}
\end{aligned}
\label{set:BD}
\end{equation}

where $\dd\Omega_3^2$ is the unit round metric on $S^3$. Its horizon $A=0$ has areal radius $R_h$, with

\begin{equation}
\mu=R_h^2+\tfrac12\tilde\alpha.
\label{set:muh}
\end{equation}

\paragraph{Dimensionless form.}\label{app:vacexterior}

The final exterior is the Boulware--Deser solution~\eqref{set:BD} with

\begin{equation}
\begin{aligned}
F_\lambda(R)&=\frac{2\mu_{\rm out}}{R^2+(R^4+2\lambda\mu_{\rm out})^{1/2}},\\
A_\lambda&=1-F_\lambda,\qquad\mu_{\rm out}=1+\tfrac12\lambda,
\end{aligned}
\label{vac:bd}
\end{equation}

which is smooth at $\lambda=0$. Its horizon $R=1$ is nonextremal.

At zero coupling, the exterior reduces to $F_0(R)=R^{-2}$. The radial derivative $A_\lambda'(1)$ can also be checked directly by differentiating the constant mass relation $R^2F_\lambda+\frac12\lambda F_\lambda^2=\mu_{\rm out}$ at fixed $\lambda$:

\begin{equation}
\begin{aligned}
0&=2RF_\lambda+(R^2+\lambda F_\lambda)F_\lambda'\\
A_\lambda'(1)&=-F_\lambda'(1)=\frac{2}{1+\lambda}>0
\end{aligned}
\label{vac:horizonderivative}
\end{equation}

The second line uses $F_\lambda(1)=1$. Thus the horizon remains nonextremal as $\lambda$ tends to zero.

\subsection{Exact constraint data}\label{app:vacdata}

As in Ref.~\cite{Suryaatmadja2026}, let $\sigma_i$ be the left-invariant one-forms on $S^3$ normalized by $\dd\sigma_1=-2\sigma_2\wedge\sigma_3$ cyclically, so that $\sum_i\sigma_i^2$ is the unit round metric. On a finite collar crossing the horizon in Painlev\'e--Gullstrand slicing, prescribe the spatial metric and extrinsic curvature

\begin{equation}
\begin{aligned}
\gamma&=\dd R^2+R^2\sum_{i=1}^3\sigma_i^2,\\
\mathcal K^i{}_j&={\rm diag}\bigl(P,H+\Delta,H+\Delta,H-2\Delta\bigr),
\qquad\Delta=\epsilon\,\Delta_0(R),
\end{aligned}
\label{vac:data}
\end{equation}

in the frame $(\dd R,R\sigma_1,R\sigma_2,R\sigma_3)$. Here $\mathcal K_{ij}=-\frac12\mathcal L_{e_0}\gamma_{ij}$ with $e_0$ the future unit normal, as in Ref.~\cite{Suryaatmadja2026}. On functions of $R$ a prime denotes $\dd/\dd R$. The profile $\Delta_0$ is supported in a compact interval strictly inside $R=1$; it is fixed below, after Eq.~\eqref{vac:bump}. The metric $\gamma$ is flat for every $\epsilon$, and the tracefree angular part of $\mathcal K$ specifies the initial time derivative of the tensor perturbation.

The free profile $\Delta_0$ specifies the angular anisotropy, while the radial eigenvalue $P$ and mean angular eigenvalue $H$ must adjust to satisfy the constraints. Keeping this adjustment exact is important: the resulting family is a family of vacuum solutions, even though its first-order perturbation is used later to detect the obstruction.

The Hamiltonian and momentum constraints of vacuum EGB gravity~\cite{ToriiShinkai2008}, the normal--normal and normal--spatial projections of the field equations, reduce exactly to

\begin{equation}
\begin{aligned}
0&=3\bigl\{P\bigl[H+\lambda(H+\Delta)^2(H-2\Delta)\bigr]+H^2-\Delta^2\bigr\}
\\
0&=3\bigl[1+\lambda(H^2-\Delta^2)\bigr]\Bigl(\frac{P-H}{R}-H'\Bigr)
+6\lambda(H+\Delta)\Delta\Bigl(\Delta'+\frac{\Delta}{R}\Bigr)
\end{aligned}
\label{vac:constraints}
\end{equation}

For a direct check, note that since $\gamma$ is flat the Gauss equation makes the Hamiltonian constraint equal to the sum of the products of pairs of eigenvalues of $\mathcal K$ plus $3\lambda$ times the product $P(H+\Delta)^2(H-2\Delta)$ of all four; expansion gives the first line. The radial momentum constraint is proportional to

\begin{equation}
\begin{aligned}
&2\Bigl(\frac{P-H-\Delta}{R}-(H+\Delta)'\Bigr)
\bigl[1+\lambda(H+\Delta)(H-2\Delta)\bigr]\\
&\qquad+\Bigl(\frac{P-H+2\Delta}{R}-(H-2\Delta)'\Bigr)
\bigl[1+\lambda(H+\Delta)^2\bigr]
\end{aligned}
\label{vac:momdirect}
\end{equation}

and expansion gives the second line. The symmetry group, generated by left $SU(2)$, right $U(1)$ and the discrete rotation that reverses $\sigma_2$ and $\sigma_3$, admits no invariant one-form on $S^3$. The angular momentum constraints therefore vanish identically, and the same symmetry excludes cross terms between the homogeneous three-spheres and the $(t,R)$ directions throughout the development.

The constraints are thus solved by the ordinary differential equations

\begin{equation}
\begin{aligned}
P&=-\frac{H^2-\Delta^2}{H+\lambda(H+\Delta)^2(H-2\Delta)},\\
H'&=\frac{P-H}{R}+\frac{2\lambda(H+\Delta)\Delta(\Delta'+\Delta/R)}
{1+\lambda(H^2-\Delta^2)}.
\end{aligned}
\label{vac:ode}
\end{equation}

The Painlev\'e--Gullstrand slicing of the stationary exterior has $\Delta=0$ and

\begin{equation}
H=\frac{F_\lambda^{1/2}}{R}\qquad P=\bigl(F_\lambda^{1/2}\bigr)'
\label{vac:pg}
\end{equation}

Fix the outer value of $H$ to this stationary value. At $\Delta=0$ the two denominators in Eq.~\eqref{vac:ode} are $H(1+\lambda H^2)$ and $1+\lambda H^2$, with positive lower bounds on the fixed collar, uniformly for small $\lambda$, so the equation has a smooth solution on the whole collar. Where $\Delta=0$,

\begin{equation}
\bigl[R^4\bigl(H^2+\tfrac12\lambda H^4\bigr)\bigr]'=0
\label{vac:mass}
\end{equation}

and since $F=R^2H^2$ there, Eq.~\eqref{set:mass} shows that the data coincide with Boulware--Deser data with $\mu=\mu_{\rm out}$ outside the support of $\Delta_0$ and with some $\mu=\mu_{\rm in}$ inside it. The absence of a term linear in the prescribed perturbation can be seen directly from

\begin{equation}
\begin{aligned}
(H+\Delta)^2(H-2\Delta)&=H^3-3H\Delta^2-2\Delta^3\\
(H+\Delta)\Delta\Bigl(\Delta'+\frac{\Delta}{R}\Bigr)&=\epsilon^2(H+\epsilon\Delta_0)\Delta_0\Bigl(\Delta_0'+\frac{\Delta_0}{R}\Bigr)
\end{aligned}
\label{vac:quadraticdata}
\end{equation}

Thus the explicit perturbation terms in Eq.~\eqref{vac:ode} begin at order $\epsilon^2$. Since the outer value of $H$ is independent of $\epsilon$, the first-order changes of $H$ and $P$ vanish. Consequently,

\begin{equation}
P=P_0+O(\epsilon^2)\qquad H=H_0+O(\epsilon^2)\qquad
\mu_{\rm in}=\mu_{\rm out}+O(\epsilon^2)
\label{vac:scalar}
\end{equation}

No condition is imposed on the inner mass. In particular, fixing the stationary exterior does not require the perturbation to have zero energy: its effect on the spherically symmetric data begins at second order and is absorbed into $\mu_{\rm in}$.

\subsection{Extension to a complete slice}\label{app:vacslice}

The collar extends to a complete two-ended slice through regions where the data coincide with those of the Boulware--Deser solution. Bend its outer end toward the static slices of the asymptotic region. Continue the slice inward, in regular null coordinates of the Boulware--Deser solution with $\mu=\mu_{\rm in}$, along a spacelike graph that crosses the Einstein--Rosen bridge to the other asymptotic end, remaining uniformly spacelike with an areal radius bounded below by a fixed positive constant. These choices persist under the small change of $\mu_{\rm in}$. At each join, the data on both sides agree with those induced from the same Boulware--Deser solution on an open overlap, so there are no surface layers. The flat form~\eqref{vac:data} is used only on the collar. Away from it, the data are induced from the corresponding Boulware--Deser spacetime on the chosen spacelike slice, and hence continue to satisfy the constraints. Completeness is obtained by the two asymptotically flat ends, not by continuing the flat collar toward the black-hole singularity.

\subsection{Hyperbolic development and parameter dependence}\label{app:vacdevelop}

Use a modified harmonic gauge~\cite{KovacsReall2020} invariant under the symmetry group, written with a fixed invariant reference connection on the homogeneous three-spheres. A formulation of the field equations is strongly hyperbolic if, after reduction to first order, its principal symbol $\mathsf M(\zeta)$ for each unit spatial covector $\zeta$ has a symmetrizer, that is, a positive matrix $\mathsf S(\zeta)$, depending smoothly on $\zeta$ and on the fields, with

\begin{equation}
\mathsf S(\zeta)\,\mathsf M(\zeta)=\mathsf M(\zeta)^\dagger\,
\mathsf S(\zeta)
\label{set:symmetrizer}
\end{equation}

This gives local well-posedness of the Cauchy problem in Sobolev spaces; Ref.~\cite{SarbachTiglio2012}, Section~3, reviews this theory. For sufficiently small $\lambda$ the full Kov\'acs--Reall formulation is strongly hyperbolic in this sense, with a common time function for the characteristic cones~\cite{KovacsReall2020}. Strong hyperbolicity holds for all gravitational polarizations, gauge modes and constraint modes, including perturbations that break the symmetry. Choose the two auxiliary tangent cones of the formulation nested strictly inside the light cone of $g$. The curvature of the data is uniformly bounded, including on the bridge, so its product with $\lambda$ is uniformly small for small $\lambda$.

The construction itself has one spatial variable $x$. The Lovelock equations are linear in the second time derivatives of the metric. Solving for them and introducing the first derivatives of the metric as variables gives a first-order system for the invariant fields $Y$,

\begin{equation}
\partial_tY=\Phi(\lambda,t,x,Y,\partial_xY)
\label{vac:nonql}
\end{equation}

which is not quasilinear. The radial principal symbol of the full formulation preserves the subspace of invariant fields, so the full symmetrizer restricts to a symmetrizer on that subspace; the undifferentiated metric components add a block with zero speed. Introduce $W=\partial_xY$ as an independent field. The prolonged system

\begin{equation}
\partial_tY=\Phi(\lambda,t,x,Y,W)\qquad
\partial_tW=\Phi_W\,\partial_xW+\Phi_YW+\Phi_x
\label{vac:prol}
\end{equation}

where subscripts on $\Phi$ denote partial derivatives, is quasilinear with principal matrix $\mathsf M={\rm diag}(0,\Phi_W)$ and a smooth positive symmetrizer. The difference $W-\partial_xY$ obeys

\begin{equation}
\partial_t(W-\partial_xY)=\Phi_Y\,(W-\partial_xY)
\label{vac:compat}
\end{equation}

so data with $W=\partial_xY$ give solutions of Eq.~\eqref{vac:nonql}. In the reduction to first order, the difference between each first-derivative variable in $Y$ and the corresponding derivative of the metric has zero time derivative, so it stays zero. The gauge conditions obey homogeneous equations by the Bianchi identity and therefore vanish, so the metric solves the vacuum EGB equations.

The choices are made in the following order: a regular background slab, then a fixed smooth profile supported in it, then a sufficiently small amplitude $\epsilon$. The width of the profile will be chosen in Sec.~\ref{app:vacshear}; once chosen, it is held fixed when differentiating with respect to $\epsilon$ and taking the large-radius limit. Cauchy stability around the stationary solution, that is, continuous dependence of the solution on its data on a fixed compact domain, gives a common backward development on the slab, and the profile may have large but fixed derivatives. On the noncompact ends, use finite propagation speed on patches or uniformly local norms, the suprema of Sobolev norms over unit balls. These accommodate the different inner and outer masses without requiring a square-integrable difference from the stationary data.

The same system gives the dependence on $\lambda$ and $\epsilon$. Set $\hat Y=(Y,W)$, so that Eq.~\eqref{vac:prol} is $\partial_t\hat Y=\mathsf M(\hat Y,\lambda)\,\partial_x\hat Y$ plus terms without derivatives of $\hat Y$. A first derivative $\partial_\nu\hat Y$ with $\nu\in\{\lambda,\epsilon\}$ satisfies the linear system

\begin{equation}
\partial_t(\partial_\nu\hat Y)=\mathsf M(\hat Y,\lambda)\,\partial_x
(\partial_\nu\hat Y)+\mathsf M_0(\hat Y,\partial_x\hat Y,\lambda)\,
\partial_\nu\hat Y+J_\nu
\label{vac:param}
\end{equation}

Here $\mathsf M_0$ collects the derivatives of the coefficients of Eq.~\eqref{vac:prol} with respect to $\hat Y$, and $J_\nu$ is the derivative of the right side of Eq.~\eqref{vac:prol} with respect to $\nu$ at fixed $\hat Y$ and $\partial_x\hat Y$; $J_\epsilon=0$ because $\epsilon$ enters only through the data. Higher mixed derivatives satisfy the same equation with sources that contain lower parameter derivatives and finitely many additional spatial derivatives. Energy estimates and difference quotients give, on every compact spatial patch $\mathcal P$ held fixed in the spatial coordinates throughout the slab,

\begin{equation}
\sup_t\bigl\|\partial_\lambda^{m_1}\partial_\epsilon^{m_2}\hat Y
\bigr\|_{H^m(\mathcal P)}\le C_{m,m_1,m_2}
\label{vac:parambound}
\end{equation}

for every fixed $m$, $m_1$ and $m_2$, provided the data are smooth enough. Each parameter derivative may require finitely many additional spatial derivatives; smoothness of the flow map at a fixed Sobolev order is not required. For sufficiently small $\epsilon$, the system remains uniformly strongly hyperbolic on the whole slab. Thus, for any prescribed finite collection of derivatives, one can choose a common parameter neighborhood on which the required estimates hold. This is the parameter regularity needed below; no uniform estimate over all derivative orders is asserted.

\subsection{Nonzero first-order shear}\label{app:vacshear}

Write the metric of the homogeneous three-spheres in the development as

\begin{equation}
q=R^2\bigl[e^{2\beta}(\sigma_1^2+\sigma_2^2)+e^{-4\beta}\sigma_3^2\bigr]
\label{vac:orbit}
\end{equation}

In the notation of Eq.~(D.1) of Ref.~\cite{Suryaatmadja2026}, the scale factors are $a_\perp=Re^\beta$ and $a_\parallel=Re^{-2\beta}$. Thus $R$ is the areal radius, with

\begin{equation}
\dd\mu_q=(Re^\beta)(Re^\beta)(Re^{-2\beta})\,\dd\mu_{\Omega_3}=R^3\,\dd\mu_{\Omega_3}
\label{vac:volumecancel}
\end{equation}

Here $\dd\mu_{\Omega_3}$ is the volume element of the unit round three-sphere. Comparing the angular eigenvalues of $\mathcal K$ in Eq.~\eqref{vac:data} with $\mathcal K=-\frac12\mathcal L_{e_0}\gamma$ gives, on the final slice,

\begin{equation}
\beta=0\qquad e_0(\beta)=-\Delta\qquad e_0(R)=-RH
\label{vac:beta}
\end{equation}

The tensor ${\rm diag}(1,1,-2)$ in the round frame is transverse and tracefree on $S^3$. Together with Eq.~\eqref{vac:scalar}, this shows that the first-order perturbation is a pure tensor mode: the metric of the orbit space and the position of the horizon do not change at first order.

Let $\beta_1=\partial_\epsilon\beta|_{\epsilon=0}$. A tensor master variable of Ref.~\cite{TakahashiSoda2010} is

\begin{equation}
Z=R\,\bigl[(R^2K)'\bigr]^{1/2}\beta_1\qquad
R^2K=\bigl(R^4+2\lambda\mu_{\rm out}\bigr)^{1/2}
\label{vac:master}
\end{equation}

where $(R^2K)'=2R^3/(R^2K)>0$. Its static potential is $A_\lambda$ times a smooth function near the nonextremal horizon, so in regular Kruskal coordinates $Z_{UV}+\mathcal V_\lambda(U,V)Z=0$ with a smooth $\mathcal V_\lambda$. The horizon is $U=0$, and the adjacent future interior has $U>0$. Write the final slice near the horizon as the graph $V=V_{\rm f}(U)$, with $V_{\rm f}'=\dd V_{\rm f}/\dd U<0$, and introduce

\begin{equation}
\tilde U=V_{\rm f}^{-1}(V)\qquad\tau=\tfrac12(U-\tilde U)\qquad
x=\tfrac12(U+\tilde U)
\label{vac:straight}
\end{equation}

The final slice is $\tau=0$, its interior part has $x>0$, and the horizon is $x=-\tau$. The equation becomes

\begin{equation}
Z_{\tau\tau}-Z_{xx}+\tilde{\mathcal V}_\lambda(\tau,x)Z=0\qquad
\tilde{\mathcal V}_\lambda=-4V_{\rm f}'(\tilde U)\,\mathcal V_\lambda
\label{vac:wave}
\end{equation}

with $\tilde{\mathcal V}_\lambda$ and its derivatives bounded uniformly near $\lambda=0$.

Choose a fixed smooth function $B$ with $0\le B\le1$, not identically zero and supported in $(\frac12,1)$, and prescribe on the final slice 

\begin{equation}
Z(0,x)=0\qquad Z_\tau(0,x)=B(x/w)\qquad c_B=\int B(y)\,\dd y>0
\label{vac:bump}
\end{equation}

with a width $w>0$ fixed below. On the final slice $\partial_\tau=N_{\rm f}\,e_0$ with a smooth positive function $N_{\rm f}$, so Eqs.~\eqref{vac:beta} and~\eqref{vac:master} give

\begin{equation}
Z_\tau=-R\,\bigl[(R^2K)'\bigr]^{1/2}N_{\rm f}\,\Delta_0
\label{vac:ztau}
\end{equation}

on the final slice. This fixes the profile $\Delta_0$ in Eq.~\eqref{vac:data}, which may depend smoothly on $\lambda$.

Set $t=-\tau$ and $\tilde Z(t,x)=-Z(-t,x)$, so that $\tilde Z$ solves the same type of equation forward in $t$ with $\tilde Z(0,x)=0$ and $\tilde Z_t(0,x)=B(x/w)$. On the triangle $0\le t\le w$, $t\le x\le2w-t$, Duhamel's formula reads

\begin{equation}
\tilde Z(t,x)=\frac12\int_{x-t}^{x+t}B(y/w)\,\dd y
-\frac12\int_0^t\int_{x-(t-t')}^{x+(t-t')}\tilde{\mathcal V}_\lambda(-t',y)\,
\tilde Z(t',y)\,\dd y\,\dd t'
\label{vac:duhamel}
\end{equation}

The first term is the solution obtained when the potential is set to zero. On the horizon $x=t$, it becomes

\begin{equation}
\left.\frac12\int_{x-t}^{x+t}B(y/w)\,\dd y\right|_{x=t}=\frac{w}{2}\int_0^{2t/w}B(s)\,\dd s
\label{vac:freehorizon}
\end{equation}

Since $B$ is supported in $(1/2,1)$, this term vanishes for $0\le t<w/4$ and equals $c_Bw/2$ at $t=w$. The second term in Eq.~\eqref{vac:duhamel} is the correction due to the potential; the following estimate shows that it cannot cancel this positive value when $w$ is sufficiently small.

Let $c_*$ bound $|\tilde{\mathcal V}_\lambda|$ on a fixed regular patch that contains the triangle, and let $Z_{\max}$ be the largest value of $|\tilde Z|$ on the triangle. The first integral in Eq.~\eqref{vac:duhamel} is at most $w$ in absolute value, and the double integral is at most $\frac12c_*w^2Z_{\max}$, so

\begin{equation}
Z_{\max}\le w+\tfrac12c_*w^2Z_{\max}
\label{vac:zmax}
\end{equation}
Choose $w$ so that the triangle lies in the patch and
$c_*w^2\le\min(1,c_B/4)$. Then
\begin{equation}
Z_{\max}\le2w\qquad\tilde Z(w,w)\ge\tfrac12c_Bw-c_*w^3\ge\tfrac14c_Bw>0
\label{vac:nonzero}
\end{equation}

For $0\le t<w/4$ the horizon point $(t,t)$ depends only on the data on $[0,2t]$, where they vanish, so $\tilde Z(t,t)=0$. The first-order anisotropy of the intrinsic metric on horizon cuts is therefore nonzero at early times and zero at late times. Along the background horizon, $R=1$, so the positive radial factor relating $Z$ to $\beta_1$ in Eq.~\eqref{vac:master} is constant. Hence $\beta_1$ has the same nontrivial time dependence: its derivative along the generators is nonzero somewhere and vanishes on a final open interval. These are the two bracketed stretches in Figure~\ref{fig:vacfamily}.

On homogeneous cuts of the horizon, Eqs.~\eqref{cut:qdot}
and~\eqref{vac:orbit} give
\begin{equation}
\chi^{+A}{}_B={\rm diag}\Bigl(\frac{\ell_+(R)}{R}+\ell_+(\beta),\
\frac{\ell_+(R)}{R}+\ell_+(\beta),\
\frac{\ell_+(R)}{R}-2\ell_+(\beta)\Bigr)
\label{vac:eigen}
\end{equation}

in the round frame. The first variations of $R$ and of the orbit-space metric vanish, so

\begin{equation}
\begin{aligned}
\chi^{+A}{}_B&=\epsilon\,\Sigma^A{}_B+O(\epsilon^2),\qquad
\theta_+=O(\epsilon^2),\\
{\rm tr}\,\Sigma&=0,\qquad\Sigma\not\equiv0,
\end{aligned}
\label{vac:shear}
\end{equation}

with $\Sigma^A{}_B=\ell_+(\beta_1)\,{\rm diag}(1,1,-2)^A{}_B$, which vanishes on a final open interval. The estimate~\eqref{vac:nonzero} holds at $\lambda=0$ as well, so these properties also hold on the Ricci-flat family used in the entropy reduction. The role of the wave calculation is to establish this nonzero first-order shear; the spacetime itself is the exact nonlinear development of the constraint data.

\subsection{Stationary future and the event horizon}\label{app:vachorizon}

Choose $R_c<1$ between the support of $\Delta_0$ and the horizon, in an open collar where the final data agree with the Boulware--Deser data. In the region where the solution is exactly Boulware--Deser, the inverse effective metrics $G_I^{ab}$ of the tensor, vector and scalar gravitational sectors are

\begin{equation}
K\,G_I^{ab}=g_{\rm orb}^{ab}+\kappa_I\,g_{\rm sph}^{ab}\qquad
\kappa_T=\frac{3K^2-2}{K^2}\qquad
\kappa_V=\frac{1}{K^2}\qquad
\kappa_S=\frac{2-K^2}{K^2}
\label{vac:effective}
\end{equation}

where, as in Ref.~\cite{Suryaatmadja2026}, $g_{\rm orb}^{ab}$ and $g_{\rm sph}^{ab}$ are the parts of $g^{ab}$ along the two-dimensional orbit space of the time and radial directions and along the round spheres, and $K=1+\lambda F_\lambda/R^2$. These are Eqs.~(3.1) and (3.3) of Ref.~\cite{Suryaatmadja2026} without a scalar field; see also Ref.~\cite{ReallTanahashiWay2014}. The radius $R_c$ is bounded below by the smallest areal radius on the fixed patch used in Eq.~\eqref{vac:duhamel}, so on $R\ge R_c$ one has $K=1+O(\lambda)$ uniformly. Hence every $\kappa_I$ is positive for small $\lambda$, and every $G_I$ is Lorentzian with the time--radial block of $g$. At $R=R_c$ the function $A_\lambda$ is negative, so

\begin{equation}
G_I^{ab}\,\partial_aR\,\partial_bR=\frac{A_\lambda}{K}<0\qquad I=T,V,S
\label{vac:dR}
\end{equation}

and $\dd R$ is timelike for every physical characteristic cone. Every future-directed characteristic therefore crosses this surface toward smaller $R$. The auxiliary metrics of the formulation are chosen with cones inside the light cone, so they have the same property. Choose a coordinate time vector that is timelike for every cone. The characteristics of the variables with zero coordinate speed follow this vector and therefore cross $R=R_c$ toward smaller $R$, although the surface may move in these coordinates. This uses the characteristic cones of the exactly stationary spherical background only, and the strong hyperbolicity used throughout is that of the formulation described above.

The direction of the characteristics is used here only at the boundary of the region whose stationary solution is to be preserved. It does not require the metric event horizon to be a causal boundary for every gravitational polarization in the nonspherical portion of the spacetime.

The local energy estimate for Eq.~\eqref{vac:prol} has no incoming flux through $R=R_c$. Uniqueness therefore identifies the future development of the final data in $R\ge R_c$ with the exact stationary exterior for arbitrarily late times. Glue it to the backward development on their open overlap. This gives a neighborhood of the horizon and of the right asymptotic end. The construction requires neither nonlinear stability of the black hole nor regularity of its late interior.

The metric has a Lorentzian two-dimensional orbit space, a positive metric on the three-spheres and no cross terms. Decompose a causal tangent vector as $X=X_{\rm orb}+X_{\rm sph}$. Then

\begin{equation}
g_{\rm orb}(X_{\rm orb},X_{\rm orb})=g(X,X)-q(X_{\rm sph},X_{\rm sph})\le0
\label{vac:causalprojection}
\end{equation}

Thus its projection to the orbit space is causal. At each point, the outer boundary of the allowed radial directions is the outgoing radial null direction. 

Trace $R=1$ backward from the stationary region along outgoing radial null curves. Solutions of this ordinary differential equation preserve their order, so points outside the traced surface reach the stationary exterior and then $\mathscr I^+_{\rm R}$, while points inside cannot cross it, and later the stationary interior prevents escape. The traced surface is therefore the corresponding part of $\partial J^-(\mathscr I^+_{\rm R})$. Compactness and smallness of $\epsilon$ exclude caustics on the developed slab. The generators are future complete because their affine parameter grows exponentially with Killing time in the nonextremal stationary region.

Smooth dependence of the null equation on parameters, together with Eq.~\eqref{vac:parambound}, gives all finite parameter derivatives of the event horizon. Extend its identification with a fixed tube to a parameter-dependent diffeomorphism of a neighborhood and pull the metrics back to the tube. The horizon is invariant under the symmetry group, and the identification is chosen equivariant, with the tube coordinate $v$ constant on the homogeneous three-spheres. Parameter derivatives of the pulled-back metrics include the derivatives of this diffeomorphism.

\subsection{Scaling at fixed coupling}\label{app:vacscaling}

Call the pulled-back metrics $\bar g_{\lambda,\epsilon}$ and set $g_{L,\epsilon}=L^2\bar g_{\lambda,\epsilon}$ with $\lambda=\tilde\alpha/L^2$. These metrics solve the EGB equations with the fixed coupling $\alpha_{(2)}$, and their curvature is $O(L^{-2})$, so the product of curvature and the fixed coupling is uniformly small for large $L$. A constant rescaling leaves the Levi-Civita connection unchanged, and an orthonormal frame $\bar e_a$ of $\bar g$ gives the orthonormal frame $e_a=L^{-1}\bar e_a$ of $g$. A geometric quantity of engineering degree $d$ is therefore $L^{-d}$ times the same quantity computed from $\bar g_{\lambda,\epsilon}$.

Equation~\eqref{vac:parambound} with $m_1=m_2=0$, together with the identification of the horizon in Sec.~\ref{app:vachorizon}, bounds any fixed number of coordinate derivatives of $\bar g_{\lambda,\epsilon}$ near the tube uniformly in $\lambda\le\lambda_0$, and with $m_1=1$ it shows that they depend continuously on $\lambda$ down to $\lambda=0$ at fixed $\epsilon$. At $\lambda=0$ the same estimate gives the $\epsilon$-derivatives that determine the first- and second-order perturbations due to the gravitational wave on a fixed dimensionless slab. These are the properties used in section~\ref{app:reduction}: taking $L\to\infty$ at fixed $\alpha_{(2)}$ isolates the lowest curvature order of the entropy, while the wave profile remains fixed in dimensionless coordinates.

\section{Reduction of local entropies at large radius}
\label{app:reduction}\label{app:coefficients}

The entropy class in section~\ref{supp:setup} permits arbitrarily high derivative order across its domain. We therefore cannot begin by assuming that the density consists only of area, intrinsic curvature and products of the null second fundamental forms. Instead, we derive a necessary condition on the leading curvature contribution of any entropy in that class.

Smoothness in the product topology gives a finite derivative order in a neighborhood of the flat planar jet. A finite Taylor expansion of the density there identifies the geometric scalars that can contribute at the first nontrivial order in inverse radius. A representative density need not be invariant under boosts; only its integral is. Its Taylor polynomial may therefore contain terms that are not boost invariant. Section~\ref{app:expansion} shows that invariance of the integrated functional removes every such term of the relevant degrees after integration over the cut. No choice of an invariant density is needed.

Throughout this section the physical coupling $\alpha_{(2)}$ is held fixed. The engineering degree $d$ of a geometric quantity is defined by its scaling as $L^{-d}$ in an orthonormal frame when all geometric lengths are enlarged by $L$. The large-radius limit is taken for fixed graphs in the dimensionless coordinates of the horizon tube. It is not an expansion uniform over every possible cut.

\subsection{Local finite order}\label{app:localorder}

The smoothness specified in section~\ref{supp:setup} implies finite derivative order near each jet. Write a density coefficient as $F(j)$ in independent jet coordinates, with the product topology. Joint continuity of its differential at $(j_0,0)$ gives neighborhoods $U$ of $j_0$ and $W$ of zero on which $|\dd F(j)[h]|<1$. Choose $W$ to restrict only finitely many coordinates. If $h$ vanishes in those coordinates, then $th\in W$ for every real $t$, so

\begin{equation}
|t|\,|\dd F(j)[h]|=|\dd F(j)[th]|<1\qquad\text{for every }t\in\mathbb R
\label{coef:unrestricted}
\end{equation}

It follows that $\dd F(j)[h]=0$ for every $j\in U$ and every such $h$. Shrink $U$ to a basic product neighborhood and enlarge the finite list to include its defining coordinates. If $j_1,j_2\in U$ agree on this list, the segment between them stays in $U$, and

\begin{equation}
F(j_2)-F(j_1)=\int_0^1\dd F\bigl((1-t)j_1+t j_2\bigr)[j_2-j_1]\,\dd t=0
\label{coef:finitecoordinates}
\end{equation}

Thus $F$ is a smooth function of finitely many jet coordinates on $U$. The order may depend on the neighborhood. Finite truncations of these coordinates are described in section~\ref{app:jets}.

For a fixed smooth compact cut, finitely many such neighborhoods cover its metric and cut jets. A fixed partition of unity on the cut gives a finite-order expression near that configuration. Integration by parts gives a smooth local first variation. In particular, graphs that agree on an open set have the same Euler--Lagrange density there.

At the flat planar jet, fix one such product neighborhood and the corresponding finite list of coordinates. Both are fixed throughout the large-radius reduction. Only finitely many coordinates are restricted by the neighborhood; higher derivatives are unrestricted. Joint smoothness in the metric and cut data is needed here, since orders chosen separately for each metric need not remain bounded as $L\to\infty$.

\subsection{Geometric jet coordinates}\label{app:jets}

Let $\mathcal C$ be a spacelike cut with null normals $\ell_\pm$ normalized by Eq.~\eqref{set:normals}. Normal exponential coordinates $(x^A,y^+,y^-)$ label the point reached by the geodesic that leaves the point $x$ of $\mathcal C$ with velocity $y^+\ell_++y^-\ell_-$. In these coordinates the Gauss lemma gives

\begin{equation}
y^+g_{++}+y^-g_{-+}=-y^-\qquad
y^+g_{+-}+y^-g_{--}=-y^+\qquad
y^+g_{+A}+y^-g_{-A}=0
\label{coef:radial}
\end{equation}
where $g_{\pm\pm}$, $g_{\pm A}$ are the components along $\partial_{y^\pm}$ and $\partial_A$. Conversely, if a metric satisfies Eq.~\eqref{coef:radial} and has $g_{AB}=q_{AB}$, $g_{\pm A}=0$, $g_{+-}=-1$ and $g_{\pm\pm}=0$ on $\mathcal C$, then $\Gamma_{M++}(y^+)^2+2\Gamma_{M+-}y^+y^-+\Gamma_{M--}(y^-)^2=0$ for every coordinate index $M$, so the curves $(x,ty^\pm)$ are geodesics and $(x,y)$ are the normal exponential coordinates of that metric; this is the radial identity used in tubular expansions~\cite{Mukhopadhyay2014}.

A boost $\ell_\pm\to e^{\pm b}\ell_\pm$ with $b$ a function on $\mathcal C$ changes $y^\pm\to e^{\mp b}y^\pm$ and $\omega_A\to\omega_A+\partial_Ab$. The one-forms

\begin{equation}
\vartheta^\pm=\dd y^\pm\pm\omega_Ay^\pm\dd x^A
\label{coef:theta}
\end{equation}

transform like $y^\pm$, because
\begin{equation}
\begin{aligned}
(\vartheta^\pm)'&=\dd(e^{\mp b}y^\pm)\pm(\omega_A+\partial_A b)e^{\mp b}y^\pm\dd x^A\\
&=e^{\mp b}\vartheta^\pm
\end{aligned}
\label{coef:boostcancel}
\end{equation}
The prime denotes the transformed one-form. The terms containing derivatives of $b$ cancel. The metric can be written as

\begin{equation}
\begin{aligned}
g={}&G_{AB}\dd x^A\dd x^B
+2\bigl(G_{+A}\vartheta^++G_{-A}\vartheta^-\bigr)\dd x^A\\
&+G_{++}(\vartheta^+)^2+2G_{+-}\vartheta^+\vartheta^-
+G_{--}(\vartheta^-)^2 .
\end{aligned}
\label{coef:metric}
\end{equation}

with $G_{AB}=q_{AB}+O(y)$, $G_{\pm A}=O(y^2)$, $G_{+-}=-1+O(y^2)$ and $G_{\pm\pm}=O(y^2)$. Since the normal connection matrix is antisymmetric in its lowered normal-frame indices, Eq.~\eqref{coef:radial} becomes $y^+G_{+A}+y^-G_{-A}=0$ together with the normal identities of Eq.~\eqref{coef:radial} for $G_{\pm\pm}$ and $G_{+-}$. The Taylor coefficients in $y$ of the fields $G$ transform as tensors under boosts, and $\omega$ is the only field that transforms inhomogeneously. At each order the conditions on the coefficients are linear, and any finite polynomial in $y$ whose coefficient fields satisfy them gives, through Eq.~\eqref{coef:metric}, a Lorentzian metric near $\mathcal C$ for which $(x,y)$ are normal exponential coordinates. The coefficients are therefore freely specifiable fields subject to linear conditions, whereas curvature components obey differential identities. Normal derivatives of any finite order are included by raising the order of the polynomial.

At a given point of $\mathcal C$ use Riemann normal coordinates of $q$ centred there, $x^Aq_{AB}(x)=x^A\delta_{AB}$. This condition is linear in the coefficient jets. We do not fix the boost gauge, because a representative density is not assumed to be boost invariant: the jets of $\omega_A$ at the point are retained. At each finite truncation order, constant boosts act linearly on the coefficients, with integer weights, and the jets of $\omega$ have weight zero. A boost by a function $b$ on the cut multiplies a coefficient field of weight $s$ by $e^{sb}$ and shifts $\omega_A$ by $\partial_Ab$; in particular it does not change the value at the point of any coefficient of weight zero in the fields $G$. These coefficients give coordinates on the corresponding finite geometric jet space. The metric at the point is fixed by the frame, and every other coordinate has a positive engineering degree $d$: it scales as $L^{-d}$ when the geometry is enlarged by a constant factor $L$. The degree and the weight satisfy

\begin{equation}
d\ge|\text{weight}|\qquad d\ge1,\quad\text{with }d=1\text{ only for }\chi^\pm_{AB}\text{ and }\omega_A
\label{coef:degree}
\end{equation}

To see this, the general solution of the radial conditions in Eq.~\eqref{coef:metric} is

\begin{equation}
\begin{aligned}
G_{+A}&=y^-\Xi_A,&G_{-A}&=-y^+\Xi_A,\\
G_{\pm\pm}&=(y^\mp)^2\,\Xi_\perp,&G_{+-}+1&=-y^+y^-\,\Xi_\perp .
\end{aligned}
\label{coef:radialsol}
\end{equation}

with polynomials $\Xi_A$ and $\Xi_\perp$ in $y$, where $\Xi_A(0)=0$ because $G_{\pm A}=O(y^2)$. Since $y^\pm$ have weights $\mp1$, the coefficient of $(y^+)^a(y^-)^c$ in $G_{AB}$ has weight $a-c$ and degree at least $a+c$; in $\Xi_A$ and $\Xi_\perp$ it has weight $a-c$ and degree at least $a+c+1$ and $a+c+2$, respectively. Derivatives along the cut raise the degree without changing the weight. The coordinates of degree one are the coefficients of $y^\pm$ in $G_{AB}$, which are $2\chi^\pm_{AB}$ with weights $\pm1$, and the components $\omega_A$ at the point, with weight zero. The jets of $q$ begin at degree two in Riemann normal coordinates. This gives Eq.~\eqref{coef:degree}.

\subsection{Taylor estimates and the scalars of low degree}\label{app:bounds}

A representative density is a scalar under diffeomorphisms; only under boosts may it change by a cut divergence. It depends on the normals through the frame components of the fields $G$, which carry boost weights, and through the jets of $\omega_A$ along the cut. No extension of the normalization off the cut enters. In the coordinates of section~\ref{app:jets}, which are adapted to the normals, these are the jets of the metric, and the smoothness of section~\ref{set:class} is understood for all of them, the jets of $\omega_A$ included. By covariance the density is therefore a function $F$ of the coordinates above that is invariant under rotations of the frame $e_A$; we normalize it by $S=\int_{\mathcal C}F\,\dd\mu_q$. A flat planar cut in Minkowski spacetime with constant normals has all coordinates zero. On the neighborhood fixed in section~\ref{app:localorder}, $F$ is a smooth function of finitely many coordinates $z_i$, with engineering degrees $d_i\ge1$, so ordinary Taylor expansion applies. Engineering degree counts powers of inverse length, not the number of factors in a Taylor monomial. For example, a term linear in curvature and a term quadratic in $\chi^\pm$ both have degree two. The parts of $F$ of degree zero, one and two are
\begin{equation}
\begin{aligned}
F^{[0]}&=F(0)\equiv c_0,\qquad
F^{[1]}(z)=\sum_{d_i=1}\partial_iF(0)z_i,\\
F^{[2]}(z)&=\sum_{d_i=2}\partial_iF(0)z_i+\frac12\sum_{d_i=d_j=1}\partial_i\partial_jF(0)z_i z_j .
\end{aligned}
\label{coef:weightedTaylor}
\end{equation}
Every other monomial of the second-order Taylor polynomial has degree at least three, and the ordinary Taylor remainder is cubic. Since every $d_i\ge1$, for each $M>0$ there are constants $C_M$ and $\Lambda_M$ such that
\begin{equation}
\begin{gathered}
\bigl|F(z)-F^{[0]}-F^{[1]}(z)-F^{[2]}(z)\bigr|\le C_M\Lambda^{-3}\\
\hbox{whenever}\quad|z_i|\le M\Lambda^{-d_i}\ \hbox{for all }i,\quad\Lambda\ge\Lambda_M .
\end{gathered}
\label{coef:T2}
\end{equation}
Equation~\eqref{coef:T2} requires only a finite Taylor expansion; no analyticity or polynomial dependence on curvature is assumed.

Rotations of the frame and constant boosts act linearly on the coordinates and preserve their degrees. Each $F^{[d]}$ is therefore invariant under rotations and is a sum of polynomials of definite boost weight,
\begin{equation}
F^{[d]}=\sum_nF^{[d]}_n,\qquad
F^{[d]}_n\longmapsto e^{nb}F^{[d]}_n\quad\hbox{under a constant boost }b .
\label{coef:weights}
\end{equation}
The individual terms need not vanish for $n\neq0$, since only the integral of the density is boost invariant.

We now list the scalars of weight zero. The coordinates of degree one are $\chi^\pm_{AB}$ and $\omega_A$. The rotation-invariant linear functions of them are combinations of $\theta_\pm$, with weights $\pm1$; hence $F^{[1]}_0=0$. At degree two, the products of two coordinates of degree one that have weight zero give the scalars $\theta_+\theta_-$, $\chi^+_{AB}\chi^{-AB}$ and $\omega_A\omega^A$. The coordinates of degree two and weight zero are the curvature of $q$, which on a three-dimensional cut is determined by its Ricci tensor, the coefficient $G^{(1,1)}_{AB}$ of $y^+y^-$ in $G_{AB}$, the value $\Xi_\perp(0)$ and the derivatives $\partial_A\omega_B$. Their scalar contractions are $R_q$, $q^{AB}G^{(1,1)}_{AB}$, $\Xi_\perp(0)$ and $D^A\omega_A$. Derivatives of $\chi^\pm$ along the cut have nonzero weight, the coefficients of $\Xi_A$ carry a single cut index, and the antisymmetric part of $\partial_A\omega_B$ has no scalar contraction. On a three-dimensional cut the alternating tensor gives no additional scalar from these fields, so the list is complete, including for functionals that depend on orientation. Thus
\begin{equation}
F^{[2]}_0=k_1R_q+k_2\theta_+\theta_-+k_3\chi^+_{AB}\chi^{-AB}
+k_4q^{AB}G^{(1,1)}_{AB}+k_5\Xi_\perp(0)+k_6\omega_A\omega^A+k_7D^A\omega_A
\label{coef:F20}
\end{equation}
with constants $k_i$. The first five terms are unchanged by a boost with an arbitrary function $b$ on the cut, because they contain no derivative along the cut of a coefficient of nonzero weight. The last two are not.

The two coefficients of the metric in Eq.~\eqref{coef:F20} are curvature components. Write $\mathcal R_{ABCD}=\mathcal R_{abcd}e_A^ae_B^be_C^ce_D^d$ for the spacetime curvature projected onto the cut, $\mathcal R_\parallel=q^{AC}q^{BD}\mathcal R_{ABCD}$, $\mathcal R_{+-}=\mathcal R_{ab}\ell_+^a\ell_-^b$ and $\mathcal R_{+-+-}=\mathcal R_{abcd}\ell_+^a\ell_-^b\ell_+^c\ell_-^d$. A direct computation of the curvature of the metric~\eqref{coef:metric} at the point gives
\begin{equation}
\Xi_\perp(0)=-\frac13\mathcal R_{+-+-},
\qquad
q^{AB}G^{(1,1)}_{AB}=2\chi^+_{AB}\chi^{-AB}-2\mathcal R_{+-}+2\mathcal R_{+-+-},
\label{coef:jetcurvature}
\end{equation}
with no dependence on $\omega_A$ or $\partial_A\omega_B$. The Gauss equation and curvature contractions give

\begin{equation}
\mathcal R=\mathcal R_\parallel-4\mathcal R_{+-}+2\mathcal R_{+-+-}
\qquad
\mathcal R_\parallel=R_q+2\theta_+\theta_--2\chi^+_{AB}\chi^{-AB}
\label{coef:gauss}
\end{equation}
so that
\begin{equation}
\mathcal R_{+-+-}=\frac12(\mathcal R-R_q)+2\mathcal R_{+-}-\theta_+\theta_-+\chi^+_{AB}\chi^{-AB}
\label{coef:nullcurvature}
\end{equation}
The five boost-invariant terms in Eq.~\eqref{coef:F20} therefore combine into

\begin{equation}
F^{[2]}_0-k_6\omega_A\omega^A-k_7D^A\omega_A=a_1R_q+a_2\theta_+\theta_-
+a_3\chi^+_{AB}\chi^{-AB}+a_4\mathcal R+a_5\mathcal R_{+-}
\label{coef:T2form}
\end{equation}

with constants $a_i$ that may depend on the fixed coupling and fixed parameters of the entropy.

\subsection{Expansion of the entropy and boost weights}\label{app:expansion}

A configuration $\mathfrak c=(g,\mathcal C,\ell_\pm)$ consists of a smooth Lorentzian metric defined near a smooth compact spacelike surface $\mathcal C$ of codimension two, together with null normals normalized by Eq.~\eqref{set:normals}. It need not satisfy any field equation. For $\Lambda>0$ let $\mathfrak c_\Lambda=(\Lambda^2g,\mathcal C,\Lambda^{-1}\ell_\pm)$. The frame, the Riemann normal coordinates and the normal exponential coordinates of $\mathfrak c_\Lambda$ are those of $\mathfrak c$ rescaled by $\Lambda$, so at each point of the cut
\begin{equation}
z_i(\mathfrak c_\Lambda)=\Lambda^{-d_i}z_i(\mathfrak c).
\label{red:jetscaling}
\end{equation}
A family of configurations is called bounded if the finitely many coordinates $z_i$ on which $F$ depends in the planar neighborhood and the areas of the cuts are bounded, uniformly on the cuts and over the family. For sufficiently large $\Lambda$ the jets of $\mathfrak c_\Lambda$ then lie in the planar neighborhood, and hence in the off-shell domain of the density, for every member of the family.

For every entropy in the class of section~\ref{set:class} and every bounded family,
\begin{equation}
S[\mathfrak c_\Lambda]=c_0\Lambda^3\mathcal A[\mathfrak c]
+\Lambda\int_{\mathcal C}\bigl(a_1R_q+a_2\theta_+\theta_-
+a_3\chi^+_{AB}\chi^{-AB}+a_4\mathcal R+a_5\mathcal R_{+-}\bigr)\dd\mu_q+O(1)
\label{red:expansion}
\end{equation}
as $\Lambda\to\infty$, uniformly on the family. The area, the integrand and the measure on the right are those of $\mathfrak c$. The constants $c_0=F(0)$ and $a_1,\dots,a_5$ depend only on the entropy and on the fixed coupling. No monotonicity is assumed here.

To prove Eq.~\eqref{red:expansion}, insert Eq.~\eqref{red:jetscaling} into the estimate~\eqref{coef:T2} and use $\dd\mu_{\Lambda^2q}=\Lambda^3\dd\mu_q$. This gives, uniformly on the family,
\begin{equation}
S[\mathfrak c_\Lambda]=\Lambda^3I_0[\mathfrak c]+\Lambda^2I_1[\mathfrak c]+\Lambda I_2[\mathfrak c]+O(1),
\qquad
I_d[\mathfrak c]=\int_{\mathcal C}F^{[d]}\bigl(z(\mathfrak c)\bigr)\dd\mu_q ,
\label{red:rawexpansion}
\end{equation}
with $I_0=c_0\mathcal A$, which does not depend on the normals. The functionals $I_d$ are integrals of polynomials in the jet coordinates and are defined for every configuration.

Let $b$ be a smooth function on $\mathcal C$ and let $\mathfrak c^b=(g,\mathcal C,e^{\pm b}\ell_\pm)$ be the boosted configuration. A boost commutes with the rescaling, and the integrated entropy is boost invariant on its domain, so $S[(\mathfrak c^b)_\Lambda]=S[\mathfrak c_\Lambda]$ for all large $\Lambda$. Applying Eq.~\eqref{red:rawexpansion} to the two configurations $\mathfrak c$ and $\mathfrak c^b$ and comparing the coefficients of $\Lambda^2$ and $\Lambda$ gives
\begin{equation}
I_d[\mathfrak c^b]=I_d[\mathfrak c],\qquad d=1,2,
\label{red:Idinvariant}
\end{equation}
for every configuration and every $b$. Here the off-shell domain is used: the rescaling is applied to an arbitrary configuration, with no relation between $\Lambda$ and the coupling.

Take $b$ constant. By Eq.~\eqref{coef:weights}, Eq.~\eqref{red:Idinvariant} reads $\sum_n(e^{nb}-1)\int_{\mathcal C}F^{[d]}_n\,\dd\mu_q=0$ for every real $b$. The functions $e^{nb}-1$ with distinct $n\neq0$ are linearly independent, so
\begin{equation}
\int_{\mathcal C}F^{[d]}_n\bigl(z(\mathfrak c)\bigr)\dd\mu_q=0\qquad(n\neq0,\ d=1,2)
\label{red:nullweights}
\end{equation}
on every configuration. Since $F^{[1]}_0=0$, this gives $I_1=0$, and $I_2=\int_{\mathcal C}F^{[2]}_0\,\dd\mu_q$ with $F^{[2]}_0$ given by Eq.~\eqref{coef:F20}.

Now take a nonconstant function $b$ and boost by $tb$ with $t$ real. The first five terms of Eq.~\eqref{coef:F20} are unchanged, $\omega_A$ becomes $\omega_A+t\partial_Ab$, and Eq.~\eqref{red:Idinvariant} gives
\begin{equation}
0=I_2[\mathfrak c^{tb}]-I_2[\mathfrak c]
=k_6\int_{\mathcal C}\bigl(2t\,\omega^AD_Ab+t^2D^AbD_Ab\bigr)\dd\mu_q
+k_7t\int_{\mathcal C}D^AD_Ab\,\dd\mu_q .
\label{red:localboost}
\end{equation}
The coefficient of $t^2$ shows that $k_6=0$. The term $k_7D^A\omega_A$ is a divergence on the compact cut and does not contribute to $I_2$. The remaining five terms combine into Eq.~\eqref{coef:T2form}, as shown in section~\ref{app:bounds}. This proves Eq.~\eqref{red:expansion}.

The terms of nonzero weight and the terms containing $\omega$ may well be present in a given representative density. The argument shows only that they integrate to zero at the two orders that matter. This is why no invariant representative has to be constructed.

\subsection{Stationary normalization}\label{app:subtract}

Write the Boulware--Deser metric of horizon radius one in ingoing coordinates, $\bar g_\lambda=-A_\lambda\dd v^2+2\dd v\,\dd R+R^2\dd\Omega_3^2$, with the normals $\bar\ell_+=\partial_v$ and $\bar\ell_-=-\partial_R$ on the round cuts of its horizon $R=1$. These metrics depend smoothly on $\lambda$ down to $\lambda=0$, so the round cuts form a bounded family, and the stationary black hole of radius $L$ at fixed coupling is the rescaled configuration with $\Lambda=L$ and $\lambda=\tilde\alpha/L^2$. On these cuts $\chi^+=0$, $R_q=6$ in units of the horizon radius, and the area is $A_3$. The Ricci tensor of $\bar g_\lambda$ vanishes at $\lambda=0$ and is therefore $O(\lambda)=O(L^{-2})$. Equation~\eqref{red:expansion} and the stationary normalization~\eqref{set:stat} give
\begin{equation}
A_3\bigl(L^3+12\alpha_{(2)}L\bigr)=c_0A_3L^3+6a_1A_3L+O(1).
\label{red:cpar}
\end{equation}
Comparing the terms of order $L^3$ and of order $L$,
\begin{equation}
c_0=1,\qquad a_1=2\alpha_{(2)} .
\label{red:a1}
\end{equation}
The stationary entropy thus fixes the area coefficient and the intrinsic-curvature coefficient. The two extrinsic-curvature terms vanish on stationary cuts, and their coefficients remain unrestricted at this stage. They must therefore be retained when testing the dynamical cuts. Only round cuts of the stationary horizons are used; the bifurcation cut plays no special role.

\subsection{The reduced functional and its necessary identities}
\label{app:S2}

The reduction leaves two undetermined constants $a_2,a_3$, depending on the proposed entropy and on the fixed physical coupling. The auxiliary functional
\begin{equation}
S_2[\psi]=\int_{\mathcal C_\psi}
\bigl(2\alpha_{(2)}R_q+a_2\theta_+\theta_-
+a_3\chi^+_{AB}\chi^{-AB}\bigr)\dd\mu_q
\label{red:S2}
\end{equation}
on the Ricci-flat family $\bar g_{0,\epsilon}$ must satisfy, for every graph $\psi$ in the horizon interval and every $|\epsilon|<\epsilon_0$,
\begin{equation}
S_2[\psi]=\frac{1}{A_3}\int_{S^3}S_{2,\rm hom}\bigl(\psi(x)\bigr)\,\dd\mu_{\Omega_3}(x),
\label{red:S2ultralocal}
\end{equation}
where $S_{2,\rm hom}(s)$ is the value of $S_2$ on the homogeneous cut $\psi\equiv s$. Consequently
\begin{equation}
\left.\partial_\epsilon^m\mathcal E^{AB}[S_2]\right|_{\epsilon=0}=0,
\qquad m=1,2 .
\label{red:S2id}
\end{equation}
Bars on the cut quantities in equation~\eqref{red:S2} are suppressed. These are necessary conditions for an exact entropy law. The functional $S_2$ is extracted from the expansion~\eqref{red:expansion}; it is not an assumption about the derivative order of $S$, nor is it asserted to approximate $S$ uniformly on arbitrary cuts.

To prove these statements, fix $\epsilon$ and a graph $\psi$ in the dimensionless coordinates of the tube, with $\bar\ell_+=\partial_v$. For $0<\lambda\le\lambda_0$ consider the cut $\mathcal C_\psi$ of $\bar g_{\lambda,\epsilon}$ and the homogeneous cuts $\psi\equiv s$ with $\min\psi\le s\le\max\psi$. Their jet coordinates are smooth functions of finitely many coordinate derivatives of $\bar g_{\lambda,\epsilon}$ at the tube, tangential and transverse, and of finitely many derivatives of $\psi$. By the uniform bounds on the family in section~\ref{supp:setup}, these configurations form a bounded family. The member $g_{L,\epsilon}=L^2\bar g_{\lambda,\epsilon}$ of Eq.~\eqref{red:family} is the rescaled configuration with $\Lambda=L$. With Eq.~\eqref{red:a1}, the expansion~\eqref{red:expansion} gives
\begin{equation}
\begin{aligned}
S[\psi]&=\mathcal A[\psi]+L\,\mathcal J_{\lambda,\epsilon}[\psi]+O(1),\\
\mathcal J_{\lambda,\epsilon}[\psi]&=\int_{\mathcal C_\psi}\bigl(2\alpha_{(2)}\bar R_q+a_2\bar\theta_+\bar\theta_-
+a_3\bar\chi^+_{AB}\bar\chi^{-AB}+a_4\bar{\mathcal R}+a_5\bar{\mathcal R}_{+-}\bigr)\dd\mu_{\bar q},
\end{aligned}
\label{red:scaledegree}
\end{equation}
where $S$ and $\mathcal A$ are evaluated on $g_{L,\epsilon}$, the bars denote quantities of $\bar g_{\lambda,\epsilon}$, and the remainder is bounded uniformly in $L$ for $\psi$ and for all the homogeneous cuts. Higher-degree terms of the entropy are contained in this remainder; they are not assumed to be absent.

Suppose that the entropy is nondecreasing on every ordered pair of cuts of every member. Then Eq.~\eqref{cut:homogeneous} holds on every member. The area satisfies the same identity exactly. Subtracting, and using Eq.~\eqref{red:scaledegree} for $\psi$ and for the homogeneous cuts,
\begin{equation}
L\left(\mathcal J_{\lambda,\epsilon}[\psi]
-\frac1{A_3}\int_{S^3}\mathcal J_{\lambda,\epsilon,\rm hom}\bigl(\psi(x)\bigr)\dd\mu_{\Omega_3}(x)\right)=O(1),
\label{red:reduced_finiteL}
\end{equation}
where $\mathcal J_{\lambda,\epsilon,\rm hom}(s)$ is the value on the homogeneous cut $\psi\equiv s$. Divide by $L$ and let $L\to\infty$ at fixed coupling, so that $\lambda=\tilde\alpha/L^2\to0$. The functional $\mathcal J_{\lambda,\epsilon}$ involves at most two derivatives of the metric at the cut, tangential and transverse, and $\bar g_{\lambda,\epsilon}$ converges to $\bar g_{0,\epsilon}$ with these derivatives, uniformly at the tube. Hence
\begin{equation}
\mathcal J_{0,\epsilon}[\psi]=\frac1{A_3}\int_{S^3}\mathcal J_{0,\epsilon,\rm hom}\bigl(\psi(x)\bigr)\dd\mu_{\Omega_3}(x).
\label{red:reduced_limit}
\end{equation}
At $\lambda=0$ the spacetime Ricci tensor vanishes for every $\epsilon$, so both Ricci terms in $\mathcal J_{0,\epsilon}$ vanish on every cut and $\mathcal J_{0,\epsilon}=S_2$. This is Eq.~\eqref{red:S2ultralocal}.

Varying $\psi$ in Eq.~\eqref{red:S2ultralocal} shows that the Euler--Lagrange density of the explicit functional $S_2$ is $A_3^{-1}S_{2,\rm hom}'(\psi(x))$ times the density of $\dd\mu_{\Omega_3}$. It is independent of all derivatives of $\psi$, so $\mathcal E^{AB}[S_2]=0$ on $\bar g_{0,\epsilon}$ for every $|\epsilon|<\epsilon_0$. The Ricci-flat family is twice continuously differentiable in $\epsilon$ together with the finitely many coordinate derivatives that enter $\mathcal E^{AB}[S_2]$, which gives Eq.~\eqref{red:S2id}.

The order of the operations matters. The identity~\eqref{cut:homogeneous} is exact at every finite $L$ and is a consequence of monotonicity on unrestricted cuts. The expansion~\eqref{red:expansion} is then applied to the two sides of this identity, which involve only the fixed graph $\psi$ and homogeneous cuts. Only values of the entropy enter; no derivative of the Taylor remainder with respect to the graph, the amplitude or the coupling is needed. The amplitude derivatives in Eq.~\eqref{red:S2id} act on the explicit functional $S_2$ of the Ricci-flat family.

\section{The shear obstruction}\label{sec:obstruction}

The reduction of Sec.~\ref{app:reduction} leaves the functional $S_2$ of Eq.~\eqref{red:S2}, with a fixed intrinsic-curvature coefficient and two unrestricted extrinsic-curvature coefficients. We now evaluate the necessary identities~\eqref{red:S2id} on the Ricci-flat family $\bar g_{0,\epsilon}$. Bars on its cut quantities will be suppressed. The calculation first eliminates $a_3$ at linear order in the wave amplitude and then identifies a tracefree term in $\mathcal E^{AB}[S_2]$ at quadratic order that the remaining $a_2$ term cannot cancel.

\subsection{The transverse null normal of a graph}\label{shear:graphnormal}

Let $e_A$ denote the tangent vectors of the constant-$v$ cuts, transported so that $[\ell_+,e_A]=0$. The tangent vectors and transverse null normal of the graph are
\begin{equation}
E_A=e_A+\psi_A\ell_+,
\qquad
(\ell_-)_\psi=\ell_-+\psi^Ae_A
                  +\frac12|D\psi|^2\ell_+.
\label{shear:normal}
\end{equation}
All quantities on the right are evaluated at $v=\psi(x)$, with
$\psi^A=q^{AB}\psi_B$ and $|D\psi|^2=q^{AB}\psi_A\psi_B$.
The two conditions
$g((\ell_-)_\psi,E_A)=0$ and $g((\ell_-)_\psi,(\ell_-)_\psi)=0$, together with $g(\ell_+, (\ell_-)_\psi)=-1$, verify the normalization directly.

The outgoing second fundamental form is unchanged by the change of tangent basis:
$(\chi^+_\psi)_{AB}=\chi^+_{AB}|_{v=\psi(x)}$.
Using Eq.~\eqref{shear:normal}, the definitions~\eqref{supp:chi} and
$\partial_vq_{AB}=2\chi^+_{AB}$ gives the transverse form
\begin{equation}
\begin{aligned}
(\chi^-_\psi)_{AB}={}&\chi^-_{AB}+\psi_{AB}
 +2\omega_{(A}\psi_{B)}+\kappa\psi_A\psi_B\\
&-2\psi_{(A}\chi^+_{B)C}\psi^C
 +\frac12|D\psi|^2\chi^+_{AB}.
\end{aligned}
\label{shear:transverseform}
\end{equation}
For example, the $\kappa\psi_A\psi_B$ term follows from
$g(\nabla_{\ell_+}\ell_-,\ell_+)=\kappa$; the two terms quadratic in the slope and linear in $\chi^+$ account for the $v$ dependence of the inverse cut metric and the final term in Eq.~\eqref{shear:normal}. Taking the trace gives
\begin{equation}
(\theta_-)_\psi=\theta_-+D^AD_A\psi+2\omega^A\psi_A
 +\left(\kappa+\frac12\theta_+\right)|D\psi|^2
 -2\chi^{+AB}\psi_A\psi_B.
\label{shear:transverseexpansion}
\end{equation}
These formulas make explicit how both allowed corrections depend on spatial derivatives of $\psi$.

We will use a simple local variational identity. Suppose a coordinate density has the form
\begin{equation}
\mathcal L
=L_0(x,\psi)+C^A(x,\psi)\partial_A\psi
 +A^{AB}(x,\psi)\partial_A\partial_B\psi
 +B^{AB}(x,\psi)\partial_A\psi\,\partial_B\psi,
\label{shear:genericdensity}
\end{equation}
with $A^{AB}$ and $B^{AB}$ symmetric. The Euler--Lagrange expression is
$\partial\mathcal L/\partial\psi
 -\partial_A(\partial\mathcal L/\partial(\partial_A\psi))
 +\partial_A\partial_B(\partial\mathcal L/\partial(\partial_A\partial_B\psi))$.
At zero slope, differentiating this expression with respect to the graph Hessian gives
\begin{equation}
\left.\frac{\partial\mathcal E}{\partial\psi_{AB}}\right|_{\psi_A=0}
=2\partial_vA^{AB}-2B^{AB}.
\label{shear:genericresponse}
\end{equation}
One factor $\partial_vA^{AB}$ comes from differentiating $\mathcal L$ with respect to its explicit $\psi$ argument, and the second comes from the two total derivatives of $A^{AB}$. The term quadratic in first derivatives contributes $-2B^{AB}$. The terms $L_0$ and $C^A\partial_A\psi$ give no Hessian dependence. When the second derivative in the density is written covariantly, the Christoffel-symbol term is absorbed into $C^A$.

\subsection{The coefficient \texorpdfstring{$a_3$}{a3}}\label{shear:firstorder}

On the stationary round background,
$\chi^+_0=0$, $\omega_0=0$ and $\chi^-_0$ is proportional to $q_0$.
The family satisfies
$\chi^+=\epsilon\Sigma+O(\epsilon^2)$,
$\theta_+=O(\epsilon^2)$ and ${\rm tr}_{q_0}\Sigma=0$.
To first order in $\epsilon$, the terms in the $a_3$ correction that contain spatial derivatives of $\psi$ are therefore
\begin{equation}
\epsilon a_3\int\sqrt{q_0}\,
\Sigma^{AB}\bigl(\psi_{AB}+\kappa\psi_A\psi_B\bigr)\,\dd^3x.
\label{shear:a3density}
\end{equation}
Here $\kappa$ has its background value. Terms containing $\omega$ or two factors of $\chi^+$ start at second order. Since $q_0$ is independent of $v$, Eq.~\eqref{shear:genericresponse} gives
\begin{equation}
\left.\partial_\epsilon\mathcal E^{AB}[S_2]\right|_{\epsilon=0}
=-2a_3\sqrt{q_0}\,(\kappa\Sigma-\dot\Sigma)^{AB},
\label{shear:a3}
\end{equation}
where a dot denotes $\partial_v$. As shown below, the intrinsic-curvature term and the product of the null expansions contribute to $\mathcal E^{AB}[S_2]$ only at second order and above.

The first identity in Eq.~\eqref{red:S2id} requires the right side of Eq.~\eqref{shear:a3} to vanish everywhere on the horizon interval under consideration. If $a_3\ne0$, then each component of $\Sigma$ obeys the homogeneous equation $\dot\Sigma=\kappa\Sigma$ along a generator. The shear vanishes on a final open interval, so uniqueness for this ordinary differential equation would force it to vanish throughout the interval. This contradicts the nonzero shear in the vacuum construction. Consequently,
\begin{equation}
a_3=0.
\label{shear:a3zero}
\end{equation}

\subsection{The product of the null expansions}\label{shear:expansion}

For $a_2\int\theta_+\theta_-\,\dd\mu_q$, Eq.~\eqref{shear:transverseexpansion} gives the coefficients in Eq.~\eqref{shear:genericdensity} as
\begin{equation}
\begin{aligned}
A^{AB}&=a_2\sqrt q\,\theta_+q^{AB},\\
B^{AB}&=a_2\sqrt q\,\theta_+
 \left[\left(\kappa+\frac12\theta_+\right)q^{AB}
                         -2\chi^{+AB}\right].
\end{aligned}
\label{shear:a2coefficients}
\end{equation}
The connection terms again contribute only to $C^A$. Using
$\partial_v\sqrt q=\theta_+\sqrt q$ and
$\partial_vq^{AB}=-2\chi^{+AB}$, we obtain
\begin{equation}
\partial_v A^{AB}
=a_2\sqrt q\left[(\dot\theta_++\theta_+^2)q^{AB}
                         -2\theta_+\chi^{+AB}\right].
\label{shear:a2derivative}
\end{equation}
The terms containing $\theta_+\chi^{+AB}$ cancel in
$2\partial_v A^{AB}-2B^{AB}$. Hence, at zero slope, the exact derivative with respect to the graph Hessian satisfies
\begin{equation}
\mathcal E^{AB}\left[a_2\int\theta_+\theta_-\,\dd\mu_q\right]X_{AB}
=-2a_2\sqrt q\left(\kappa\theta_+-\dot\theta_+
                              -\frac12\theta_+^2\right){\rm tr}_qX.
\label{shear:a2}
\end{equation}
Thus $\mathcal E^{AB}$ for this term is proportional to $q^{AB}$ even when the shear is nonzero.

Choose $X$ tracefree with respect to the background metric $q_0$ and keep it fixed while differentiating in $\epsilon$. Since $q=q_0+O(\epsilon)$,
${\rm tr}_qX=O(\epsilon)$. Since $\theta_+$ and $\dot\theta_+$ are $O(\epsilon^2)$, Eq.~\eqref{shear:a2} is then $O(\epsilon^3)$. The unrestricted coefficient $a_2$ cannot contribute to the tracefree second-order identity.

\subsection{The intrinsic-curvature term}\label{shear:curvature}

It remains to evaluate the intrinsic-curvature term, whose coefficient is fixed by stationary entropy. At zero slope the Hessian variation holds $q$ and $\chi^+$ fixed. Tracing Eq.~\eqref{cut:ricci} therefore gives
\begin{equation}
\delta_XR_q=2\chi^{+AB}X_{AB}-2\theta_+{\rm tr}_qX.
\label{shear:scalarresponse}
\end{equation}
Equations~\eqref{cut:JMeuler}, \eqref{cut:ricci} and~\eqref{shear:scalarresponse} imply
\begin{equation}
\begin{aligned}
\mathcal E^{AB}\left[2\alpha_{(2)}\int R_q\,\dd\mu_q\right]X_{AB}
=-8\alpha_{(2)}\sqrt q\biggl[&
 (\chi^{+2})^{AB}X_{AB}-\theta_+\chi^{+AB}X_{AB}\\
&+\frac12\left(\theta_+^2-|\chi^+|^2\right){\rm tr}_qX\biggr],
\end{aligned}
\label{shear:curvatureresponse}
\end{equation}
where $(\chi^{+2})_{AB}=\chi^+_{AC}\chi^{+C}{}_B$ and
$|\chi^+|^2=\chi^+_{AB}\chi^{+AB}$.
This contribution to $\mathcal E^{AB}[S_2]$ is quadratic in $\chi^+$ and therefore has no term linear in $\epsilon$.

Substitute $\chi^+=\epsilon\Sigma+O(\epsilon^2)$ and
$\theta_+=O(\epsilon^2)$, and take $q_0$-tracefree $X$.
The second term in the brackets is $O(\epsilon^3)$, as is the trace term. Together with Eqs.~\eqref{shear:a3zero} and~\eqref{shear:a2}, the surviving quadratic term is
\begin{equation}
\left.\frac12\partial_\epsilon^2
\mathcal E^{AB}[S_2]X_{AB}\right|_{\epsilon=0}
=-8\alpha_{(2)}\sqrt{q_0}\,(\Sigma^2)^{\TF}_{AB}X^{AB}.
\label{shear:final}
\end{equation}
Here $(\Sigma^2)_{AB}=\Sigma_{AC}\Sigma^C{}_B$, and all contractions and the tracefree projection in this equation use $q_0$.

For a nonzero tracefree symmetric $3\times3$ matrix,
$(\Sigma^2)^{\TF}$ cannot vanish. To see this, diagonalize $\Sigma$ in a $q_0$-orthonormal frame, with eigenvalues $s_1,s_2,s_3$. If its square were proportional to the identity, then
$s_1^2=s_2^2=s_3^2$. Three numbers with the same nonzero absolute value cannot sum to zero, whereas tracefreeness requires $s_1+s_2+s_3=0$. Thus $\Sigma^2$ can be proportional to the identity only if $\Sigma=0$.

At a point where $\Sigma\ne0$, choose
$X_{AB}=(\Sigma^2)^{\TF}_{AB}$, evaluated at that point. The right side of Eq.~\eqref{shear:final} becomes
\begin{equation}
-8\alpha_{(2)}\sqrt{q_0}\,
\bigl|(\Sigma^2)^{\TF}\bigr|_{q_0}^{\,2}<0.
\label{shear:nonzero}
\end{equation}
It is nonzero, contradicting the second identity in Eq.~\eqref{red:S2id}. The sign in Eq.~\eqref{shear:nonzero} concerns the derivative with respect to the graph Hessian; the actual entropy decrease follows separately from the all-cuts argument.

The dimension of the cut matters at this step. A tracefree symmetric $2\times2$ matrix has eigenvalues $s,-s$, so its square is proportional to the identity. The tracefree contraction used in Eq.~\eqref{shear:final} would then vanish. For three-dimensional cuts, every nonzero first-order shear supplies the required obstruction.

\subsection{A finite-radius violation for the exact entropy}\label{shear:exactviolation}

Let $S$ be an entropy in the class of section~\ref{set:class} with the stationary normalization; monotonicity is not assumed. Its constants $a_2$ and $a_3$ are defined by Eq.~\eqref{red:expansion}. If $a_3\ne0$, the right side of Eq.~\eqref{shear:a3} is nonzero somewhere, since otherwise $\Sigma$ would vanish identically by the argument that follows that equation. If $a_3=0$, the right side of Eq.~\eqref{shear:final} is nonzero wherever $\Sigma\ne0$. On the stationary background $\mathcal E^{AB}[S_2]$ vanishes, and it is twice continuously differentiable in $\epsilon$. In either case $\mathcal E^{AB}[S_2]$ is therefore nonzero at some point of the Ricci-flat horizon for every sufficiently small nonzero amplitude, so the Euler--Lagrange density of $S_2$ depends on the Hessian of $\psi$. Fix such an amplitude. Equation~\eqref{red:S2ultralocal}, if valid for all graphs, would make that density independent of all derivatives of $\psi$. It therefore fails for some graph $\psi$:
\begin{equation}
\delta_2\equiv S_2[\psi]-\frac{1}{A_3}\int_{S^3}S_{2,\rm hom}\bigl(\psi(x)\bigr)\,\dd\mu_{\Omega_3}(x)\neq0 .
\label{shear:defect}
\end{equation}
The expansion~\eqref{red:scaledegree} holds whether or not the entropy is monotone. Apply it to this graph and to the homogeneous cuts of the exact solution $g_{L,\epsilon}$. The area terms cancel, because the area satisfies Eq.~\eqref{cut:homogeneous} exactly, and the limit $\lambda\to0$ is taken as in section~\ref{app:S2}. This gives
\begin{equation}
S[\psi]-\frac{1}{A_3}\int_{S^3}S_{\rm hom}\bigl(\psi(x)\bigr)\,\dd\mu_{\Omega_3}(x)
=L\bigl(\delta_2+o(1)\bigr)+O(1)\qquad(L\to\infty).
\label{shear:finiteL}
\end{equation}
For every sufficiently large finite $L$, the exact entropy thus violates Eq.~\eqref{cut:homogeneous} on the fixed smooth graph $\psi$, and so it cannot be nondecreasing on all ordered cuts of that member. The proof in section~\ref{app:positivity} locates the failure. If the Euler--Lagrange density of $S$ were nonnegative on the rounded graphs~\eqref{cut:smoothminmax} for every value of the rounding parameter, the bounds~\eqref{cut:minmaxbounds} would give Eq.~\eqref{cut:homogeneous}. Hence that density is negative at some point of one of these graphs, and section~\ref{app:pair} produces an ordered pair of smooth compact cuts with strictly decreasing entropy.

The large-radius expansion was applied to values of the entropy on a fixed graph and on homogeneous cuts, where the dimensionless graph data are bounded, and the amplitude expansion to the explicit functional $S_2$. Neither approximates the entropy uniformly on arbitrarily bent cuts. Unrestricted cuts enter only through the positivity argument of Sec.~\ref{app:allcuts}, which is applied to the exact functional at finite $L$. Thus no perturbative error estimate is extrapolated to unbounded spatial derivatives of $\psi$.

\section{Discussion}\label{sec:discussion}

The obstruction comes from combining stationary normalization with the freedom to change the cut. Monotonicity on all ordered cuts forces the local entropy rate to depend only on the position of the cut along each generator. The stationary black-hole entropy, however, fixes an intrinsic-curvature contribution whose rate responds to bending when the horizon shears. Smooth local corrections cannot remove this response: the contraction of the two null second fundamental forms is excluded at first order in amplitude, and the product of the expansions cannot cancel the tracefree second-order term.

The role of the large-radius limit is to test a necessary identity for the exact entropy. It does not replace EGB gravity by Einstein gravity or truncate the proposed entropy to two derivatives. Smoothness at the planar jet and local finite order isolate the leading curvature contribution, while higher-degree terms vanish after the appropriate rescaling. The contradiction on the limiting Ricci-flat family then implies failure at some finite radius and amplitude. An ordered pair of cuts with lower entropy on the later cut follows from the all-cuts argument for the exact functional.

The result concerns smooth metric event horizons of fixed topology. The construction uses vacuum gravitational radiation, and the relevant interval and stationary future admit a strongly hyperbolic formulation. Thus neither a merger, a change in cut topology nor a loss of local well-posedness is needed. The selected horizon is defined by the metric causal structure; the argument does not identify it with a causal boundary for every gravitational polarization.

The scope of the entropy class remains essential. The density must be smooth on the stated invariant off-shell domain, including the flat planar jet, and the jet data at every point of every cut in the horizon interval must belong to that domain. Entropies singular at the planar data, functionals using additional or nonlocal information, and laws defined only for restricted families of cuts are outside the result. A law along a prescribed foliation does not impose the same restrictions as a law between every ordered pair of cuts.

The obstruction is compatible with the linearized second law~\cite{Wall2015}: its surviving term is quadratic in wave amplitude. It is also compatible with laws formulated to effective-field-theory accuracy and with derivative bounds on a chosen horizon foliation~\cite{HollandsKovacsReall2022,DaviesReall2024,DaviesReall2023}. Weak spacetime curvature alone does not enforce those cut-derivative bounds. The argument uses the stronger demand of an exact inequality on unrestricted cuts.

The large-radius scaling suggests an extension to higher-dimensional Lovelock theories with an Einstein term and a positive Gauss--Bonnet coupling, because higher Lovelock contributions to stationary entropy occur at higher curvature order. Establishing that extension would require a vacuum construction and entropy reduction in the relevant dimension. Theories without a Gauss--Bonnet term require a separate analysis. No obstruction of this generality can apply to all higher-curvature theories: nonlinear second-law results for $f(\mathcal R)$ gravity hold when $f'(\mathcal R)>0$, subject to the null energy condition and the future boundary conditions specified in Ref.~\cite{BhattacharyyaEtAl2017}.

\bibliographystyle{unsrtnat}
\bibliography{reference}
\end{document}